\documentclass{aa}
\usepackage{natbib}
\usepackage{graphicx}
\usepackage{txfonts}
\usepackage{hyperref}
\usepackage{url}
\usepackage{xcolor}
\hypersetup{
colorlinks,
linkcolor={red!80!black},
citecolor={blue!80!black},
urlcolor={blue!80!black}
}

\def\cm3{cm$^{-3}$}
\def\kms{km~s$^{-1}$}

\def\msun{M$_{\odot}$}

\def\two{\ts {\,\sc ii}}

\def\beq{\begin{equation}}
\def\eeq{\end{equation}}

\def\lesssim{\mathrel{\hbox{\rlap{\hbox{\lower4pt\hbox{$\sim$}}}\hbox{$<$}}}}
\def\gtrsim{\mathrel{\hbox{\rlap{\hbox{\lower4pt\hbox{$\sim$}}}\hbox{$>$}}}}

\def\two{{\,\sc ii}}

\def\v1d{{\sc v1d}}

\def\cmfgen{{\sc cmfgen}}
\def\heracles{{\sc heracles}}

\newcommand{\iso}[2]{\ensuremath{^{#1}\rm{#2}}}

\def\aj{AJ}

\def\apj{ApJ}
\def\apjs{ApJS}
\def\apjl{ApJL}
\def\aap{A\&A}

\def\mnras{MNRAS}
\def\nat{Nature}

\def\jqsrt{JQSRT}

\def\nifs{\iso{56}Ni}
\def\cofs{\iso{56}Co}

\begin{document}

   \title{Influence of macroclumping on type II supernova light curves}
   \titlerunning{Macroclumping and type II SN radiation}

\author{Luc Dessart\inst{\ref{inst1}}
  \and
   Edouard Audit\inst{\ref{inst2}}
  }

\institute{
Unidad Mixta Internacional Franco-Chilena de Astronom\'ia (CNRS, UMI 3386),
    Departamento de Astronom\'ia, Universidad de Chile,
    Camino El Observatorio 1515, Las Condes, Santiago, Chile\label{inst1}
\and
    Maison de la Simulation, CEA, CNRS, Universit\'e Paris-Sud, UVSQ,
    Universit\'e Paris-Saclay, 91191, Gif-sur-Yvette, France.\label{inst2}
  }

   \date{Received; accepted}

  \abstract{Core-collapse supernova (SN) ejecta are probably structured on both small and large scales, with greater deviations from spherical symmetry nearer the explosion site. Here, we present 2D and 3D gray radiation-hydrodynamics simulations of type II SN light curves from red (RSG) and blue supergiant (BSG) star explosions to investigate the impact on SN observables of inhomogeneities in density or composition, with a characteristic scale set to a few percent of the local radius. Clumping is found to hasten the release of stored radiation, boosting the early time luminosity and shortening the photospheric phase. Around the photosphere, radiation leaks between the clumps where the photon mean free path is greater.  Since radiation is stored uniformly in volume, a greater clumping can increase this leakage by storing more and more mass into smaller and denser clumps containing less and less radiation energy. An inhomogeneous medium in which different regions recombine at different temperatures can also impact the light curve. Clumping can thus be a source of diversity in SN brightness. Clumping may lead to a systematic underestimate of ejecta masses from light curve modeling, although a significant offset seems to require a large density contrast of a few tens between clumps and interclump medium.}

   \maketitle

\section{Introduction}

   The ejecta mass of core-collapse supernovae (SNe)  is a fundamental parameter characterizing these events \citep{heger_fate_03}. This is in part because numerous quantities inferred from observations scale in one way or another with the ejecta mass, such as the characteristic ejecta expansion rate or the photon diffusion time \citep{arnett_80}. Inferring the ejecta mass is also essential for estimating the progenitor mass, the nucleosynthetic yields, and building a physically consistent picture of core-collapse SN explosions (see e.g. \citealt{sukhbold_ccsn_16}).

   Paradoxically, estimating the ejecta mass, and by extension the progenitor mass, is very challenging. An origin for this difficulty may lie in the fact that the SN material holds little internal energy, which is instead all stored in radiation (produced initially by the shock but also continuously through radioactive decay). Mass is only a source of opacity, trapping the radiation, controlling its rate of escape and thus producing the resulting bolometric light curve. However, the shock-deposited energy per unit ejecta mass varies with ejecta depth. For example, the shock passage through a low-density extended envelope produces an obvious luminous burst in some type IIb SNe, which allows for an estimate of the associated mass \citep{nomoto_93j_93,podsiadlowski_93j_93,woosley_94_93j,bersten_etal_12_11dh,d18_ext_ccsn}. In contrast, the shock passage through a dense and massive He core may produce a feeble luminosity signature in type II SNe \citep{d19_sn2p}, making the inference of its mass a very delicate matter. In GRB/SNe, the ejecta masses inferred from observations around the time of bolometric maximum do not typically agree with those inferred from late time observations \citep{maeda_98bw_03,dessart_98bw_17}. A large mass must be present at low velocity, but being somewhat ``dark'', it is harder to constrain. Hence, mass inferences in different SN types, using early or late time constraints, are subject to complex uncertainties.

     In SNe powered primarily by \nifs\ decay, ejecta masses are inferred using various incarnations of a diffusion model, using assumptions such as homogeneity, fixed opacity, fixed ionization, and a negligible contribution from shock-deposited energy (see e.g. \citealt{arnett_82,chatzopoulos_lc_12}). In radiation hydrodynamics models for these ejecta, spherical symmetry, gray opacity and various levels of mixing are used (see e.g. \citealt{bersten_etal_12_11dh}). In explosions of red-supergiant (RSG) stars, the ejecta mass inferences are generally based on semi-analytical models or radiation hydrodynamics simulations (see e.g. \citealt{litvinova_sn2p_85}, \citealt{popov_93}, \citealt{utrobin_07_99em}). The ejecta is assumed spherically symmetric and smooth, and the gas is treated in local thermodynamic equilibrium (LTE). This implies that the gas ionization is set by the Saha equation.

    In reality, core-collapse SN ejecta are complex environments. The progenitor stars may not explode in a vacuum but instead in a dense and confined environment \citep{yaron_13fs_17}. The explosion is likely asymmetric on all scales, as evidenced by light echoes \citep{rest_casA_echo_11}, nebular-phase spectra \citep{fransson_chevalier_89,jerkstrand_04et_12}, late-time integral-field spectroscopic observations \citep{kjaer_87a_10}, late-time radio observations \citep{abellan_87A_17}, or spectro-polarimetric observations \citep{leonard_04dj_06}. From the theoretical point of view, these departures from spherical symmetry may arise from a variety of causes, including Rayleigh-Taylor instabilities, post-shock neutrino-driven convection, the standing-accretion-shock instability, or the effect of progenitor rotation \citep{muller_87A_91,fryxell_mueller_arnett_91,wongwathanarat_13_3d}. These give rise to asymmetries on a wide range of scales (from tens of percent to a few percent of the local radius). Smaller scale structures may also exist but are not resolved by current multi-D hydrodynamical simulations.

   It is therefore of interest to explore the impact of such complicated ejecta properties on SN radiation and quantify the impact they may have on the inferences we make from more simplistic assumptions. One such complication is the 3D inhomogeneous structure of core-collapse SN ejecta and its impact on SN observables. For example, microclumping has an effect on type II SN light curves and spectra \citep{d18_fcl}. Microclumping takes the form of density inhomogeneities that are optically thin, meaning that their scale is shorter that the typical photon mean free path. By boosting the recombination rate, microclumping hastens the recession of the photospheric layers, increases the radiation leakage from the ejecta, boosts the luminosity, shortens the rise time to maximum in BSG star explosions, and leads to a shorter photospheric phase in a type II SN. By reducing the electron density above the photosphere, it also leads to a reduction in the H$\alpha$ line strength.

   The present study investigates the influence of macroclumping on type II SN radiation properties. In contrast to microclumping, it corresponds to density inhomogeneities that are large compared to the photon mean path. Hence, macroclumping can influence the transport and escape of radiation from a SN ejecta. Here, we will consider macroclumps with a size of a few percent of the local radius,  associated with local variations in either density or  composition. Because it is not at present possible to conduct nonLTE as well as multi-D time-dependent radiative transfer, the combined influence of microclumping (which requires a solution to the statistical-equilibrium equations) and macroclumping (which requires multi-D radiative transfer) cannot be assessed.  A conclusion of this study is, however, that the two effects act in the same direction, and that assuming spherical symmetry and a smooth homogeneous medium leads to an underestimate of the ejecta mass.

In the next section, we present our numerical approach. Using gray radiation hydrodynamics in 1D, 2D, and 3D, we explore the influence of macroclumping on the SN radiation during the photospheric phase for a RSG explosion model (Section~\ref{sect_ref}). We consider various levels of density contrast (Section~\ref{sect_clump_var}) and composition (Section~\ref{sect_he}) between the clumps and the interclump medium. We also explore the influence of the progenitor radius with the case of a BSG star explosion model (Section~\ref{sect_bsg}).  Finally, we quantify the underestimate in ejecta mass that results from assuming a smooth ejecta. In Section~\ref{sect_conc}, we present our conclusions.

\section{Numerical approach}
\label{sect_setup}

\subsection{Hydrodynamics and microphysics}
\label{sect_hydro_micro}

   We have used the Eulerian multi-dimensional radiation-hydrodynamics code \heracles\ \citep{gonzalez_heracles_07,vaytet_mg_11} to perform 1-D, 2-D, and 3-D simulations of type II SNe. The code treats the hydrodynamics using a standard second order Godunov scheme. All simulations employ a gray radiation transport solver, which is based here on the M1 moment model \citep{m1_model}. As discussed in \citet{gonzalez_heracles_07}, the M1 model is well suited for the study of radiation transport in a structured medium. It captures well the shadowing  effect of a high density clump as well as the propagation speed of radiation in a transparent medium. Hence, it handles adequately the different transport properties from the optically-thick to the optically-thin layers.

 Because we start from SN ejecta that are already in homologous expansion, there is no shock on the grid. Dynamical effects are negligible and the ejecta material evolves ballistically. The gas and the radiation are in equilibrium at large optical depth and deviate modestly from each other through and above the photosphere.  The need for a multi-group treatment of the radiative transfer is therefore not crucial for the computation of the bolometric light curve, so the assumption of gray transport is adequate.

We assume a uniform H-rich composition and treat the gas as ideal, with a mean atomic weight $\mu=$\,1.35 and $\gamma=$\,5/3. For simplicity, we use a simple prescription for the opacity. Since we assume a plasma at the solar composition (thus dominated by H and He), the opacity at high temperature is well described by the Rosseland mean value $\kappa_{\rm high}=$\,0.34\,cm$^2$\,g$^{-1}$. At low temperature, we adopt the low value $\kappa_{\rm low}=$\,0.001\,cm$^2$\,g$^{-1}$. As in \citet{khatami_kasen_lc_19}, we find it convenient to use an analytical form for the temperature dependence of the opacity (we ignore any explicit dependence of the mass absorption coefficient on density, but the inverse mean-free path depends explicitly on $\rho$) with
\begin{equation}
\kappa(T) = \kappa_{\rm low}+ \Big(\frac{\kappa_{\rm high} - \kappa_{\rm low}}{2}\Big)  \,\, \Big( 1 + \frac{2}{\pi} \arctan \big( \frac{T-T_{\rm ion}}{\Delta T_{\rm ion}} \big) \Big)   \,\,\, ,
\label{eq_kappa}
\end{equation}
where $T$ is the gas temperature, $T_{\rm ion}$ is a representative recombination temperature for the gas
(e.g., which is about 6000\,K for H-rich material at representative SN ejecta densities), and $\Delta T_{\rm ion}$ is the range over which the plasma opacity transitions from $\kappa_{\rm low}$ to  $\kappa_{\rm high}$ as the temperature is raised from below to above $T_{\rm ion}$ (this transition typically occurs over a narrow temperature range, so we set  $\Delta T_{\rm ion}$ to 200\,K). We assume that this gray opacity is split between a scattering component $\kappa_{\rm sca}$  and an absorption component $\kappa_{\rm abs}$, with an assumed albedo $\kappa_{\rm sca} / (\kappa_{\rm sca} + \kappa_{\rm abs})$ fixed at a value of 0.9. This is rough but adequate for a type II SN (see, e.g., \citealt{D15_2n}, as well as appendix~\ref{appendix_albedo} for a more extended discussion).

Adopting  an ideal gas equation of state ignores the impact of changes in excitation and ionization on the  pressure, the temperature, or the energy of the gas. The thermal energy of the gas is, however, a small fraction of the total ejecta energy, which is dominated by radiation and kinetic energies. This choice allows a quick determination of the thermodynamic properties analytically, saving time and avoiding numerical issues with interpolation between pre-computed table values. Because the focus of the present study is to compare between 1D, 2D, and 3D simulations with and without macroclumping and for a given SN ejecta, these simplifications are not a  concern, provided we use (and we do) the same choices for all simulations.

We use spherical coordinates $R$, $\theta$ (and $\mu = \cos \theta$), $\phi$, with $n_R$, $n_\theta$, and $n_\phi$ zones in each direction. The lowest resolution corresponds to $n_R=480$, $n_\theta=48$ (in 2D and 3D) and $n_\phi=48$ (in 3D). Higher resolutions (used only in 1D and 2D) use two and three times as many zones $n_R$ and $n_\theta$. A higher resolution is needed when the simulation starts at a young SN age, hence when the ejecta has not yet expanded to a large radius. In 2D and 3D, we simulate wedges placed arbitrarily along $\phi$ but centered in latitude along the equatorial plane (i.e., $\theta=\pi/2$). The angular wedge in $\theta$ and $\phi$ extends over 20$^\circ$ so that a sufficient number of clumps can be used to fill the grid (see Section~\ref{sect_cl_init}).

\subsection{The reference smooth ejecta structure}

The starting conditions for the radius $R$, velocity $V$, density $\rho$, and temperature $T$ of the SN ejecta are prescribed analytically. The advantage is flexibility. The specification of the density versus velocity follows the approach of \citet{chugai_hv_07}.
The ejecta density distribution $\rho(V)$ is given by
\begin{equation}
    \rho(V)  = \frac{\rho_0}{1 + (V/V_0)^k}   \label{eq_rho}
\end{equation}
where $\rho_0$ and $V_0$ are constrained by the adopted ejecta kinetic energy
$E_{\rm kin}$, the ejecta mass $M_{\rm ej}$, and the density exponent $k$ through
\begin{equation}
M_{\rm ej} = 4 \pi \rho_0 (V_0 t)^3 C_m   \,\,   ;   \,\,
E_{\rm kin} = \frac{1}{2}\frac{C_e}{C_m} M_{\rm ej} V_0^2 \,\,  ,  \label{eq_rho_0}
\end{equation}
and where
\begin{equation}
C_m =  \frac{\pi}{k\sin(3\pi/k)}       \,\,   ;   \,\,
C_e   =  \frac{\pi}{k\sin(5\pi/k)} \,\,  .
\end{equation}

   \begin{figure*}
\includegraphics[width=0.60\hsize]{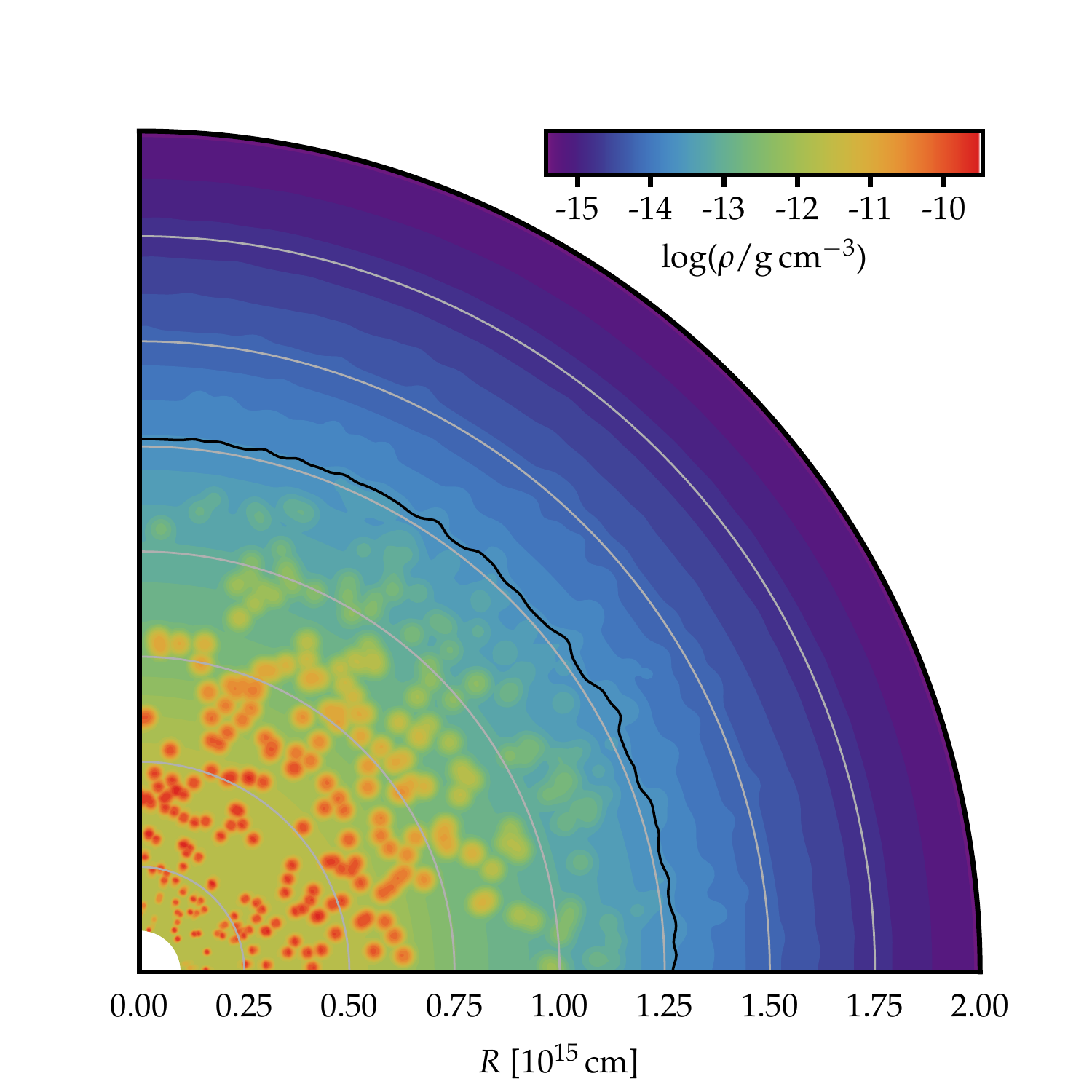}
\includegraphics[width=0.40\hsize]{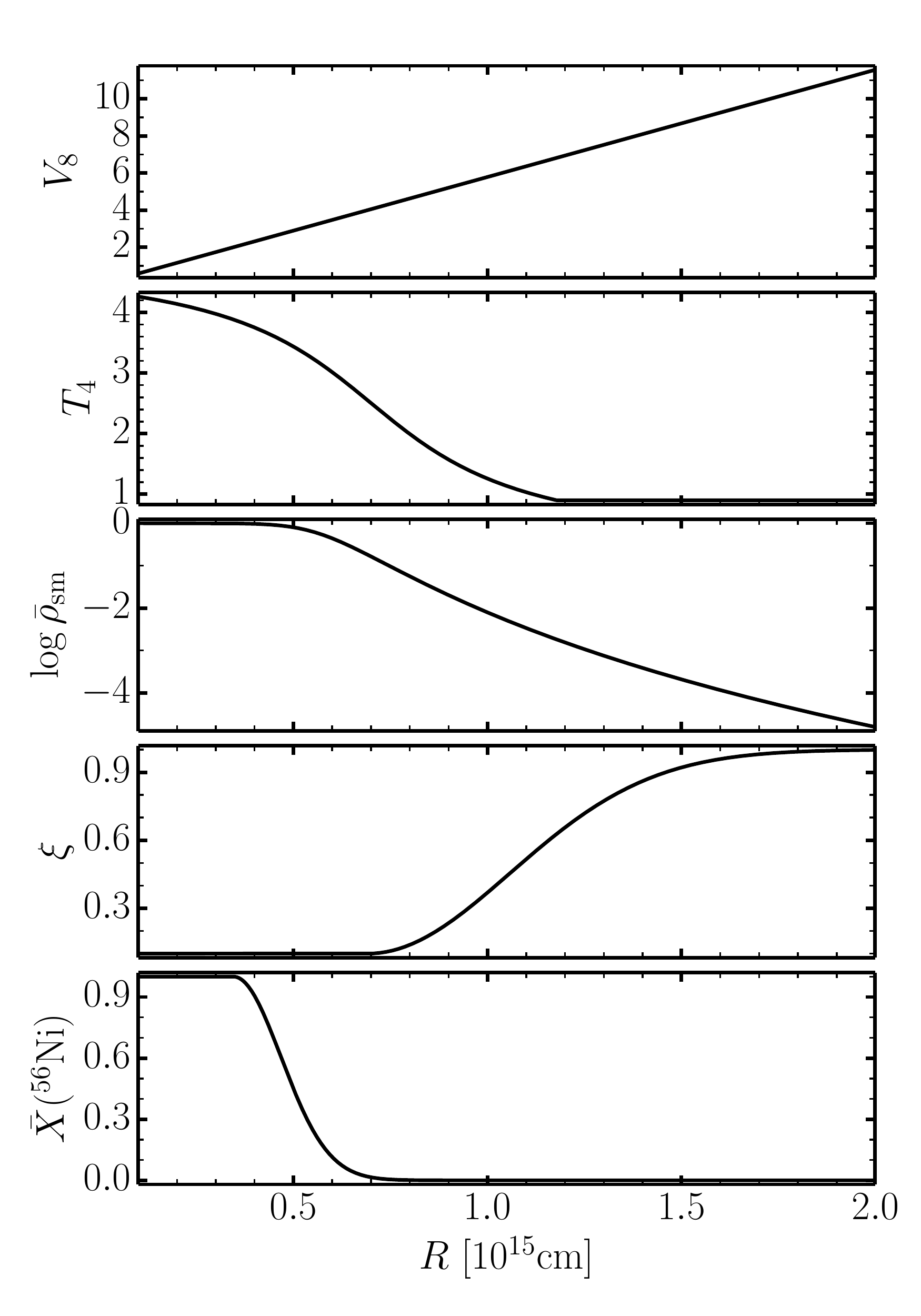}
   \caption{Two dimensional density structure (left) for a clumped model set-up over a 90$^\circ$ wedge (in our \heracles\ simulations, this wedge straddles the equatorial plane at $\theta=\pi/2$ and thus extends from $\theta=\pi/4$ to $\theta=3\pi/4$). The black line corresponds to the photosphere. Quantities other than density vary only with radius initially. The plot at right shows the velocity in units of 1000\,\kms\ ($V_8$), the temperature in units of 10$^4$\,K ($T_4$), the normalized smooth density in the log, the radial variation of clumping ($\xi$), and the normalized \nifs\ distribution $\bar{X}(^{56}{\rm Ni})$ versus radius $R$. The model corresponds to the case of a 20\,d old type II-P SN ejecta with the following model parameters: $M_{\rm ej}=$\,12\,\msun, $E_{\rm kin}=1.2 \times 10^{51}$\,erg, $k=9$, $T_{\rm high}=50,000$\,K, $T_{\rm low}=100$\,K, $R_T=7\times10^{14}$\,cm, $\Delta R_T=3\times10^{14}$\,cm,  $\sigma_{\rm cl}=0.02$, $\xi_0=0.1$, $V_{\rm cl}=4000$\,\kms, $\Delta V_{\rm cl}=3000$\,\kms, $M(^{56}{\rm Ni}) = 0.05$\,\msun, $V_{\rm Ni}=2000$\,\kms, and $\Delta V_{\rm Ni} = 1000$\,\kms.
\label{fig_init}
 }
\end{figure*}

The grid, which is Eulerian, must cover initially the space that the SN ejecta will occupy over the whole simulated evolution. The choice of grid is dictated by the impact it has on the Courant time (especially relevant when doing multi-D) or the number of zones needed to resolve the ejecta at all times. Thus, for practical reasons, the simulation is started at a SN age sufficiently large that the inner ejecta has expanded to a significant radial scale. Starting at a SN age of days to weeks after explosion, we can adopt a minimum ejecta radius between 10$^{13}$ and 10$^{14}$\,cm. For the outer radius, we must make sure that it is large enough to encompass all ejecta regions that trap radiation energy during the high brightness phase. For the ejecta properties considered here, we found that a maximum grid radius $R_{\rm max}$ of $4 \times 10^{15}$\,cm was suitable. A large fraction of the ejecta leaves the grid during the simulation but with this choice of $R_{\rm max}$, the escaping material is always optically thin and thus no longer influences the trapping or the diffusion of SN radiation (some photons interact with optically-thick lines at large velocities, which we neglect, but these interactions influence the spectral properties and not the bolometric luminosity).

The ejecta is in homologous expansion. Because the Eulerian grid extends to a large maximum radius and because the presence of a pre-SN wind is ignored, the outer velocity may become unrealistically large. Hence, the outer velocity is forced to slowly level off at $V_{\rm max}=$\,60,000\,\kms\ if it exceeds $V_{\rm lim}=$\,40,000\,\kms, following the expression :
\begin{equation}
 V(R) = V_{\rm lim} + (V_{\rm max}-V_{\rm lim}) \, ( 1 - R_{\rm lim}/R )^{0.85} \,\,\, {\rm if } \, \, \, V >  V_{\rm lim} \, ,
\end{equation}
where $R_{\rm lim} = V_{\rm lim} t$ and $t$ is the elapsed time since explosion. This feature is used for simulations of BSG explosions, which are started at an earlier time (see below). This non-homologous expansion at large velocity has no impact since it concerns only the optically thin regions of the ejecta (which quickly advect out of the grid).

In type II SNe, the shock deposited energy plays an essential part in the resulting bolometric light curve so the initial temperature structure matters. We use the following expression (similar to the opacity formulation above; Eq.~\ref{eq_kappa})
\begin{equation}
 T = T_{\rm low}+ \Big(\frac{T_{\rm high} - T_{\rm low}}{2}\Big)  \,\, \Big( 1 + \frac{2}{\pi} \arctan \big( \frac{R_T - R}{\Delta R_T} \big) \Big)   \,\,\, ,
\end{equation}
where $T_{\rm low}$ and $T_{\rm high}$ are the ejecta temperatures far from the radius $R_T$ and where $\Delta R_T$ controls the scale over which $T$
varies between $T_{\rm low}$ and $T_{\rm high}$. This expression is useful since one can mimic the presence of a temperature jump (e.g., across a recombination front) or adjust the temperature gradient as desired. In practice, using previous simulations for type II SNe as a guide \citep{d13_sn2p,d19_sn2pec}, \heracles\ was run in 1-D and low resolution until we obtained a bolometric light curve that approximately resembled that of a standard type II-Plateau or the type II-pec SN\,1987A. One can switch between the two light curve morphologies by raising $T_{\rm high}$ (SNe II-P) or lowering $T_{\rm high}$ (SNe II-pec) in the initial model.\footnote{This difference arises from the greater cooling from expansion that affects the explosions from more compact stars like BSGs relative to RSGs. This temperature difference means that type II-P SN ejecta hold more radiative energy at $10-20$\,d after explosion than SNe II-pec, which brighten to a delayed maximum because of radioactive decay heating. Consequently, SNe II-P are much more luminous early on than SNe II-pec; see for example Section~5.1 and Figs.~11--12 of \citet{dessart_11_wr}.}

In addition, all simulations treat the radioactive decay of \nifs\ and \cofs. The code can treat both local and nonlocal energy deposition. However, for the present simulations of type II SN ejecta prior to $100-200$\,d, we can assume that the decay power is deposited locally. This requires following one species across the simulation. The adopted initial profile for \nifs\ is of the form
\begin{equation}
     X(^{56}{\rm Ni}) = X(^{56}{\rm Ni}_0) \exp(-Y^2)  \,\, {\rm with } \,\, Y = \frac{V -  V_{\rm Ni}}{\Delta V_{\rm Ni}}  \, ; \,  V \geq V_{\rm Ni} \,
 \end{equation}
and $X(^{56}{\rm Ni}) = X(^{56}{\rm Ni}_0)$  for $V < V_{\rm Ni}$. Here, $X(^{56}{\rm Ni}_0)$ is the inner ejecta mass fraction of \nifs, which is constant until $V_{\rm Ni}$ and drops exponentially beyond with a characteristic scale $\Delta V_{\rm Ni}$. It is set so that the total (spherical equivalent ejecta) mass matches a desired value (irrespective of clumping; see below).

The above expressions can be used to set the boundary conditions analytically for $V$, $\rho$, and $T$. For the SN ejecta, we use an inflow inner boundary. $V_{\rm ib}$ is given from the inner boundary radius $R_{\rm ib}$ at post-explosion time $t + t_0$ as $R_{\rm ib}/(t+t_0)$ ($t_0$ is the SN age at the start of the \heracles\ simulation).  For convenience, the internal energy was set to be constant across the inner boundary. The boundary density is determined by
 \begin{equation}
\rho_{\rm ib} = \frac{\rho_0}{1 + (V_{\rm ib}/V_0)^k} \, .
\end{equation}
The SN age is incremented at each time step in order to compute the decay power and update the inner boundary condition for the velocity and density. The SN age is given as the initial SN age plus the elapsed time since the start of the simulation. In addition, $\rho_{\rm ib}$ at post-explosion time $t+t_0$ is directly determined after updating $\rho_0$ at each time step (i.e., set through the constraint that $\rho_0 t^3$ is constant in time; this comes from the constraint of mass conservation, as seen also in Eq.~\ref{eq_rho_0}). Alternatively, we have also used a reflecting inner boundary condition (there is then no inflow of material). In this case, the results for the bolometric light curve are identical (this occurs because the mass injected on the grid contains a negligible amount of radiation energy).

For the outer boundary,  we adopt a constant internal energy through the boundary, a velocity set by homology, and a density fall-off with a power law of exponent six at all times (in practice, it should evolve with time, being nine initially (this is our choice for the value of $k$ in our simulations; see Table~\ref{tab_grid}) and decreasing as the velocity declines, but this is irrelevant given the supersonic outflow speed at the outer boundary). For the radiation, we assume a reflecting inner boundary (zero flux) and a free flow outer boundary.

\subsection{Treatment of macroclumping}
\label{sect_cl_init}

We simulate the effect of macroclumping on SN radiation by adjusting the smooth density profile given by Eq.~\ref{eq_rho}. We first impose a radial variation of the magnitude of clumping using the function
\begin{equation}
     \xi(V) = 1 + (\xi_0 -1) \exp(-Y^2)  \,\, {\rm with } \,\, Y = \frac{V -  V_{\rm cl}}{\Delta V_{\rm cl}}  \,  {\rm if }\,  V \geq V_{\rm cl} \,    \label{eq_xi}
 \end{equation}
and $\xi=\xi_0$ if $V < V_{\rm cl}$. With this choice, we can impose clumping in the inner (lower velocity) ejecta regions while leaving the outer regions untouched (same density as in the smooth case given by Eq.~\ref{eq_rho}).

In our clumped models, the parameter $\xi_0$ controls the density contrast between the clump (or interclump) medium with the corresponding smooth model. When initializing a 2D/3D clumped density structure we first set the interclump density as
\begin{equation}
\rho_{\rm inter-cl}(R,\mu,\phi) = \rho_{\rm sm}(R,\mu,\phi) \, \xi(R)  \label{eq_rho_inter_cl}\, .
\end{equation}
This defines the ``background" density. We then randomly distribute clumps between $R_{\rm min}$ and $R_{\rm max}$, $\mu_{\rm min}$ and $\mu_{\rm max}$, $\phi_{\rm min}$ and $\phi_{\rm max}$. At $(R,\mu,\phi)$, the density associated with a clump at location $(R_{\rm cl},\mu_{\rm cl},\phi_{\rm cl})$ is given by
\begin{equation}
\rho_{\rm cl}(R,\mu,\phi) = \frac{(1 - \xi(R))}{\xi(R)} \rho_{\rm sm}(R,\mu,\phi)   \exp \Big(-\frac{d^2_{\rm cl}}{\sigma^2_{\rm cl} R^2_{\rm cl}} \Big)  \,\,\,\, ,    \label{eq_rho_cl}
\end{equation}
where $d_{\rm cl}$ is the distance between the clump center $(R_{\rm cl},\mu_{\rm cl},\phi_{\rm cl})$ and the location $(R,\mu,\phi)$.
The characteristic scale of a clump is $\sigma_{\rm cl} R_{\rm cl}$ to reflect spherical expansion. Our choice also implies that all clumps have the same spatial extent at a given $R$.

In reality, macroclumps may have a distribution of sizes, perhaps growing continuously from being much smaller than a photon mean free path (microclumps) to being much larger. Unlike for the treatment of microclumping, which considers that clumps are surrounded by vacuum, our macroclumps are surrounded by inter-clump material of finite density. Our  parameterization is numerically convenient but others are possible. Because of numerical limitations, we adopt relatively large clumps so that a high resolution is not needed. We therefore do not consider a distribution of clump sizes, nor consider how the radiation transport may be affected as clumps are increased from a microscopic to a macroscopic scale.

To initialize a simulation, we keep adding such clumps until the cumulative mass of the clumps plus the interclump medium equals that for the corresponding smooth ejecta model. Equations~\ref{eq_rho_inter_cl} and \ref{eq_rho_cl} indicate that for a clump at $R$ the ratio $\rho_{\rm cl}/\rho_{\rm inter-cl}$ is equal to $(1 - \xi(R))/\xi^2(R)$, which is at most $(1-\xi_0)/\xi_0^2$ -- this ratio is unity at large velocities relative to $V_{\rm cl}$ because we impose that clumping eventually dies out as we progress from the inner to the outer ejecta layers. Since the ejecta mass $\int \rho dv$ is unchanged in the presence of clumping, the clump density goes as the inverse of the volume filling factor when $\rho_{\rm cl}/\rho_{\rm inter-cl} >> 1$, as for microclumping.

In all models, the material composition is uniform throughout the ejecta, except for \nifs. The adopted \nifs\ distribution is uniform in angle but varies with radius. In other words, at a given radius or velocity,  the clump and the interclump media have the same composition (this holds in all simulations apart from Section~\ref{sect_he}). For a given choice of mixing and clumping properties, the mass fraction of \nifs\ is renormalized so that the volume integral $\int dv \rho X(^{56}{\rm Ni})$ is equal to a prescribed value (independent of the adopted clumping). With this procedure, the impact of macroclumping on a type II SN light curve can be gauged for a given ejecta mass, kinetic energy, and \nifs\ mass. This treatment of clumping leaves the bulk ejecta properties unchanged -- it merely redistributes the density over the ejecta volume under specified geometric constraints.

For most of the 2D and 3D simulations presented here, the angular wedge extends over 20$^\circ$ in $\theta$ and $\phi$. The characteristic scale of clumps is typically $0.02 R$, so about 17 can fit in the lateral direction, and a few hundred clumps are used to fill the grid. There is thus no need to use a larger angular extent.

\begin{table*}
\begin{center}
\caption{Summary of initial model parameters (see Section~\ref{sect_grid_pres} for discussion).}
\begin{tabular}{
l@{\hspace{4mm}}c@{\hspace{4mm}}c@{\hspace{4mm}}
c@{\hspace{4mm}}c@{\hspace{4mm}}c@{\hspace{4mm}}
c@{\hspace{4mm}}c@{\hspace{4mm}}c@{\hspace{4mm}}
c@{\hspace{4mm}}c@{\hspace{4mm}}
c@{\hspace{4mm}}c@{\hspace{4mm}}c@{\hspace{4mm}}
}
\hline
       Model &  $R_{\rm min}$ &  $R_{\rm max}$ &   $M_{\rm ej}$ &  $E_{\rm kin}$ &           $k$ &           Age &  $T_{\rm high}$ &  $T_{\rm low}$ &         $R_T$ &  $\Delta R_T$ &  $M(^{56}$Ni) &   $\Delta V_{\rm Ni}$ &   $V_{\rm Ni}$  \\
             & \multicolumn{2}{c}{[10$^{15}$\,cm]}  & [\msun] & [$10^{51}$\,erg] & & [d] & [kK] &  [kK] & \multicolumn{2}{c}{[10$^{15}$\,cm]} & [\msun] & \multicolumn{2}{c}{[1000\,\kms]} \\
\hline
        2P &  0.1 & 4.0 & 12 & 1.2 & 9 & 20 & 50 & 0.1  & 0.7 & 0.3 & 0.050  & 1.0  & 2.0  \\
\hline
      2pec &  0.1 & 4.0 & 13 & 1.2 & 9 & 11 & 60 & 0.1  & 0.3 & 0.1 & 0.078  & 0.5  & 2.0   \\
\hline
\label{tab_grid}
\end{tabular}
\end{center}
\end{table*}

\subsection{Set of simulations}
\label{sect_grid_pres}

   We focus on ejecta conditions typical of red-supergiant (RSG) and blue-supergiant (BSG)  progenitors, producing two sets of models called ``2P'' and ``2pec''. The main characteristic distinguishing the two sets is the initial temperature. In the 2pec set, the temperature is low initially so that the SN brightness increases with time because of the contribution from \nifs\ decay. This case corresponds to events like SN\,1987A. In the 2P set, the temperature is high initially so that the brightness is high early on and decreases with time as the ejecta releases its stored radiation energy. This corresponds to standard type II-P SNe. In this case, the decay of \nifs\ merely lengthens the high brightness phase.

   We ran simulations in 1D, 2D, and 3D, with a resolution that is larger for the 2pec set compared to the 2P set. The clumping magnitude is such that $\xi_0$ varies between 1 (smooth ejecta) and 0.1 (maximum density contrast of 90). The radial variation of clumping varies between models but is such that clumping is greater in the inner ejecta and progresses towards unity at the largest velocities (see Eq.~\ref{eq_xi}). Unless otherwise stated, and as explained in Section~\ref{sect_hydro_micro}, the opacity parameters are the same in all simulations and such that $\kappa_{\rm high}=$0.34\,cm$^2$\,g$^{-1}$, $\kappa_{\rm low}=$\,0.001\,cm$^2$\,g$^{-1}$, $T_{\rm ion}=$\,6000\,K, and $\Delta T_{\rm ion}=$\,200\,K. For \nifs, the adopted mass is 0.078\,\msun\ for the 2pec set and 0.05\,\msun\ for the 2P set (with the exception of models discussed in Section~\ref{sect_he}). This choice is arbitrary, except for the 2pec models in which \nifs\ is essential for producing a high, SN-like, luminosity. The same level of \nifs\ mixing is used for all models within a set (i.e., the set 2P or 2pec; see Table~\ref{tab_grid}).

   \begin{figure}
   \begin{center}
\includegraphics[width=\hsize]{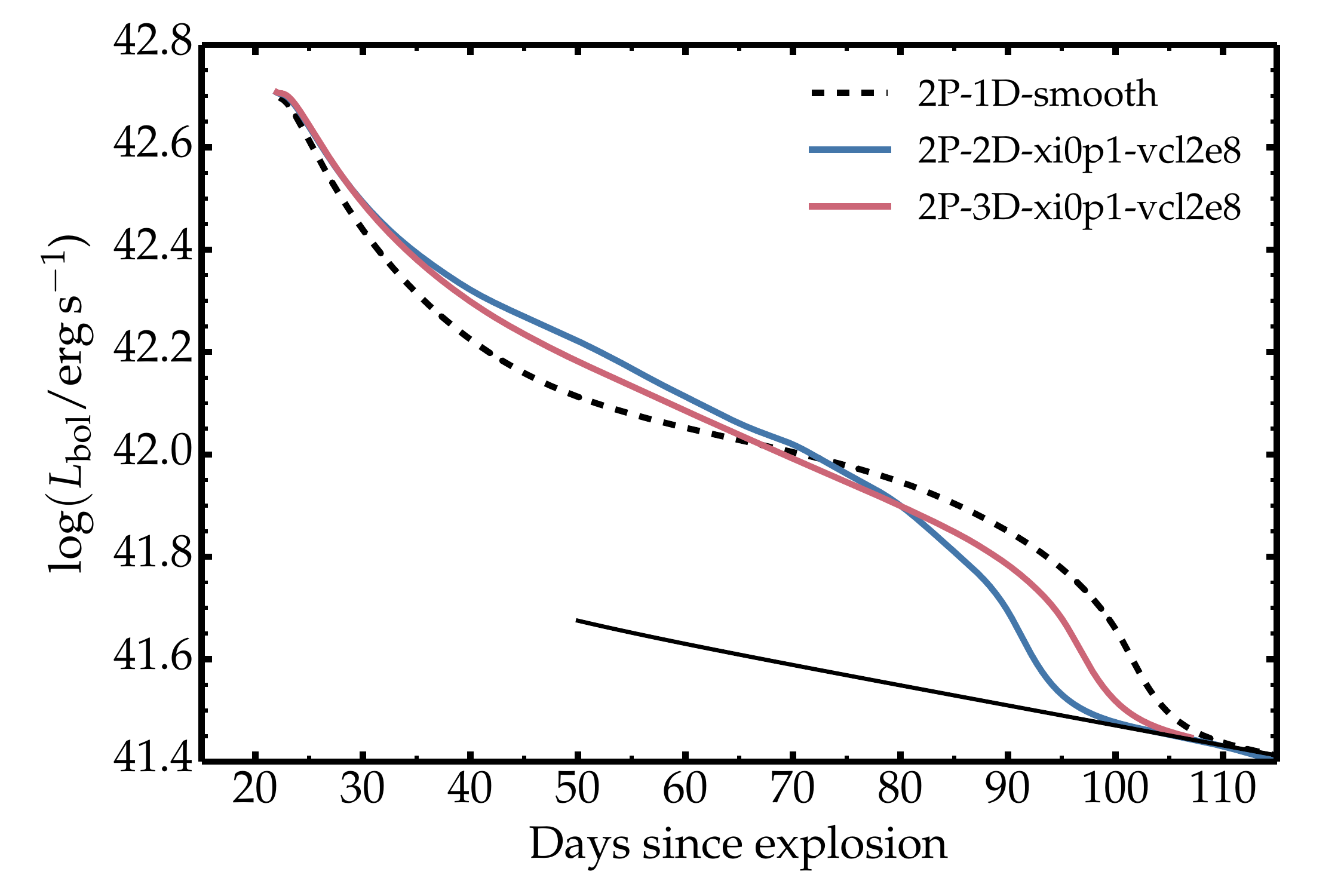}
\end{center}
   \caption{Bolometric light curves for a set of 2P models with the same ejecta properties but assuming spherical symmetry (1D and smooth; dashed line), axial symmetry (2D; clumps have a torus shape), and no symmetry (3D; clumps are spheres). The clumping parameters are $\xi_0=$\,0.1, $V_{\rm cl}=$\,2000\,\kms, $\Delta V_{\rm cl}=$\,2000\,\kms, and $\sigma_{\rm cl}=$\,0.02. The solid black line corresponds to the decay power.}
   \label{fig_lbol_2P}
\end{figure}

It is not clear whether there is a tight correlation between the profiles of \nifs\ and clumping. Both stem from fluid instabilities. The \nifs\ ``fingers" may stretch in velocity space further than the region of high clumping. Hence, the grid of simulations presented here use two different distributions for \nifs\ and clumping, and with characteristics that we allow to vary to cover a range of possibilities.

While 3D simulations have greater consistency, there is a great benefit in performing 2D simulations. They are computationally cheaper, allowing one to cover a large parameter space, and they also capture the main features of clumped ejecta on the SN radiation. Computationally expensive 3D simulations only provide a slight quantitative offset with respect to 2D counterparts. Hence, numerous simulations were performed in 2D and only four in 3D (each costs about 90,000 CPU hours; see Section~\ref{sect_ref} and \ref{sect_mej}).

Table~\ref{tab_grid} presents a summary of the model parameters used for the grid of models discussed in the following sections. Figure~\ref{fig_init} gives an illustration for one setup over a 90$^\circ$ wedge. In that case, the spatial extent of the clumped regions was enlarged to better reveal the properties of the clumps.

  \begin{figure*}[h]
   \begin{center}
     \includegraphics[width=0.33\hsize]{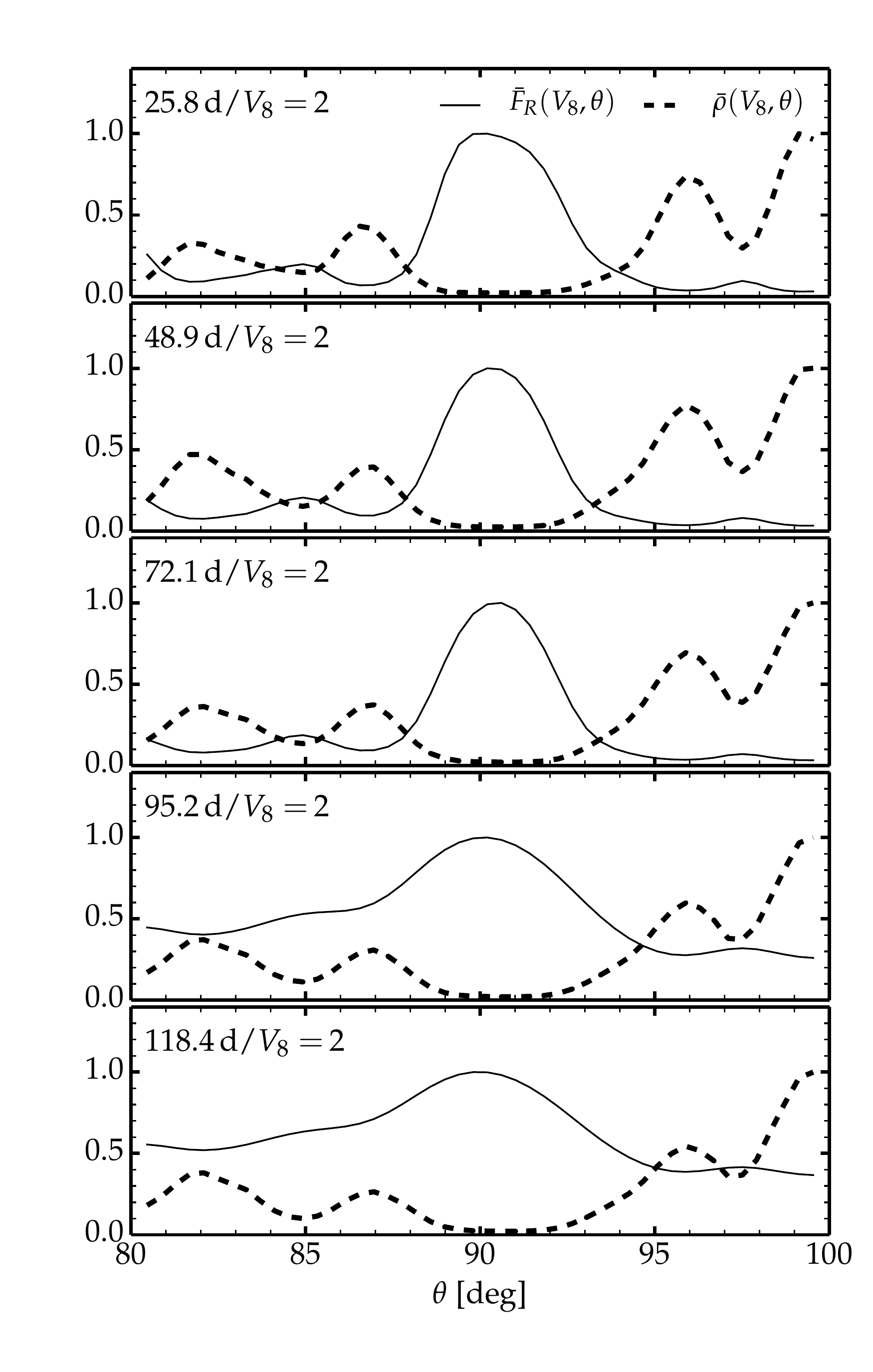}
     \includegraphics[width=0.33\hsize]{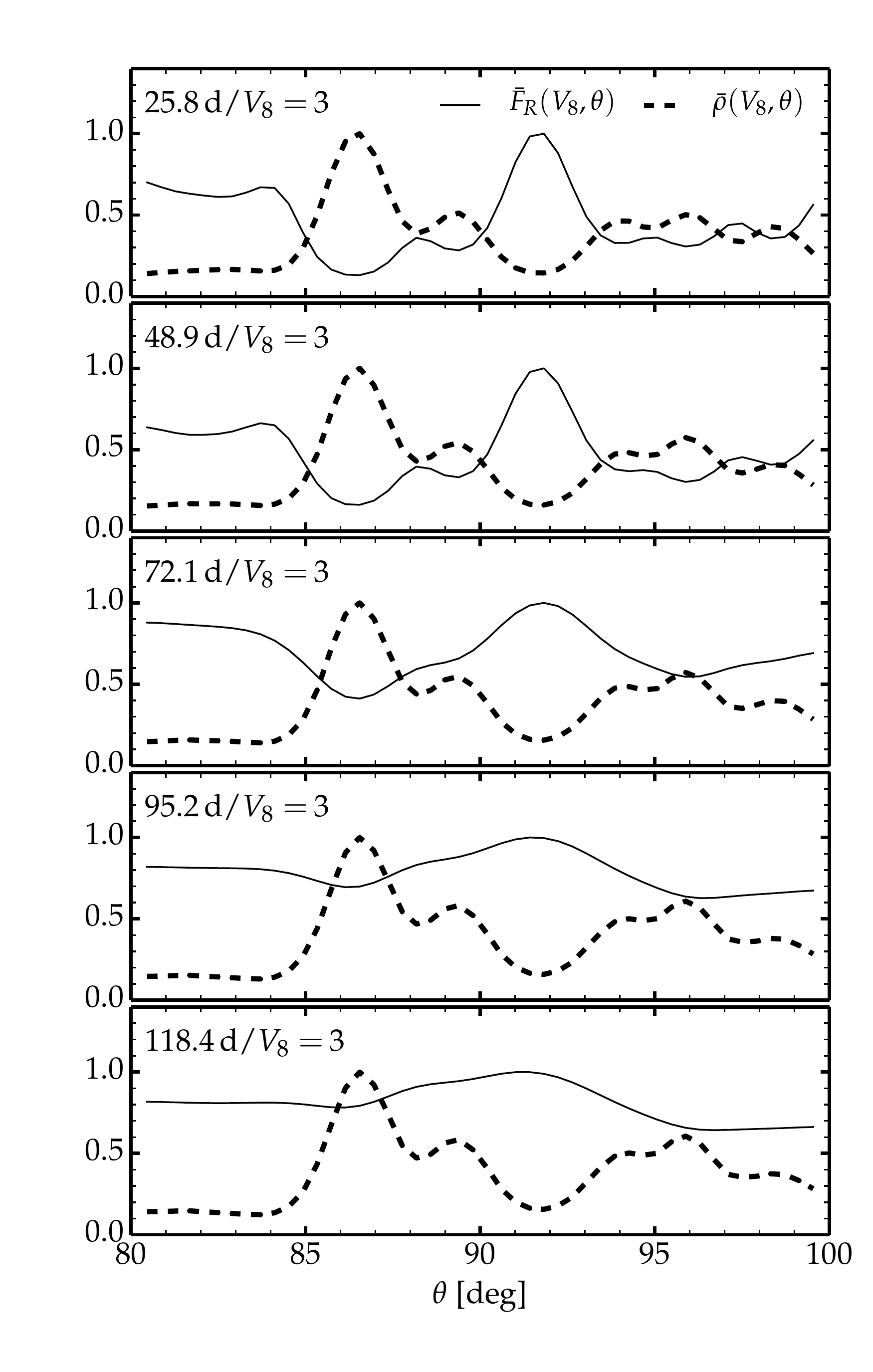}
     \includegraphics[width=0.33\hsize]{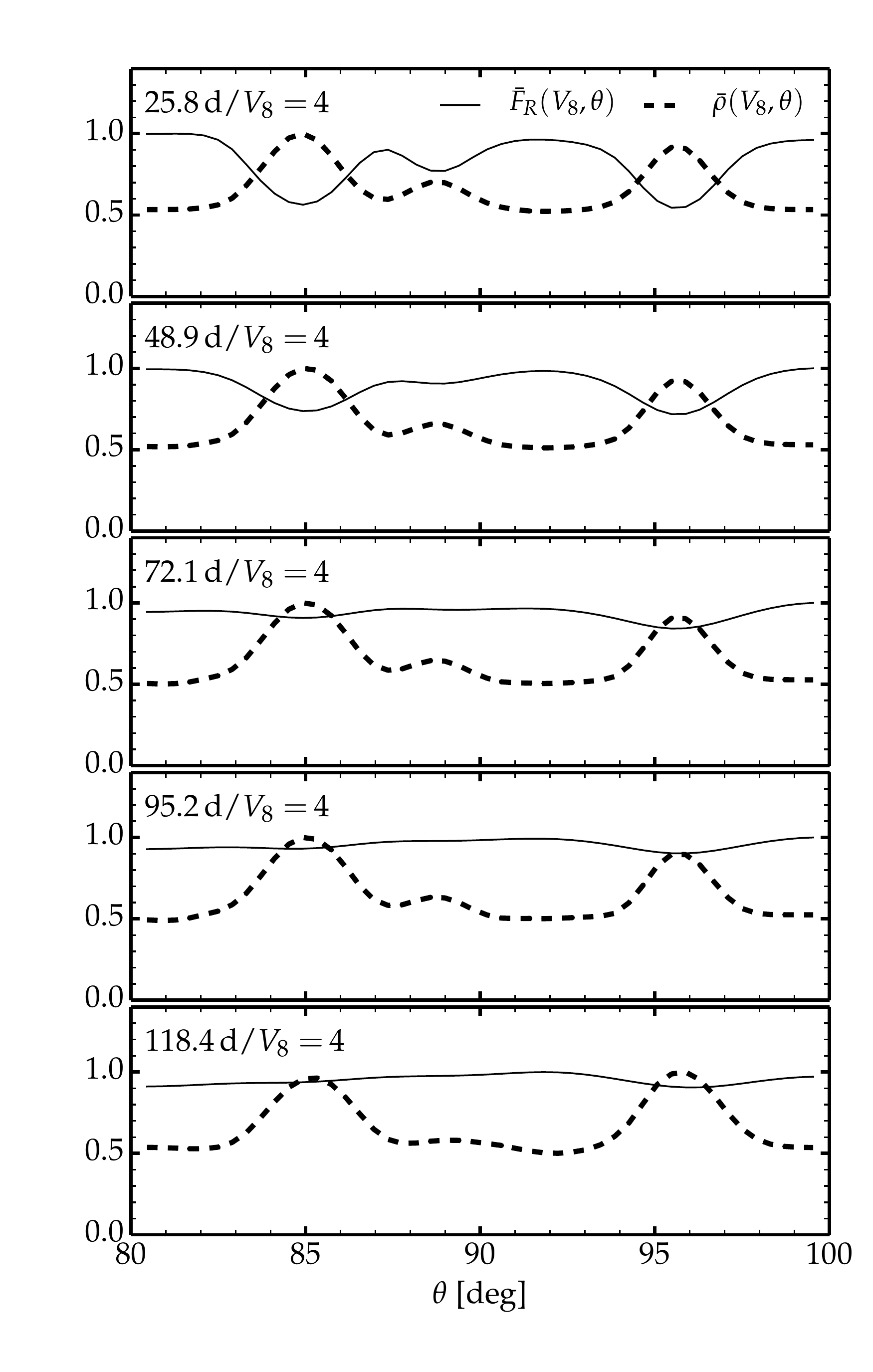}
   \end{center}
   \caption{Evolution of the normalized radial radiative flux $\bar{F}_r(V_8,\theta)$ (solid) and mass density $\bar{\rho}(V_8,\theta)$ (dashed) at three different velocities (the label $V_8$ gives the corresponding ejecta shell velocity in units of 1000\,\kms) for the 2D simulation (model 2P-2D-xi0p1-vcl2e8) shown in Fig.~\ref{fig_lbol_2P}.
   \label{fig_fr_rho}
   }
   \end{figure*}

  \begin{figure}[h]
   \begin{center}
     \includegraphics[width=\hsize]{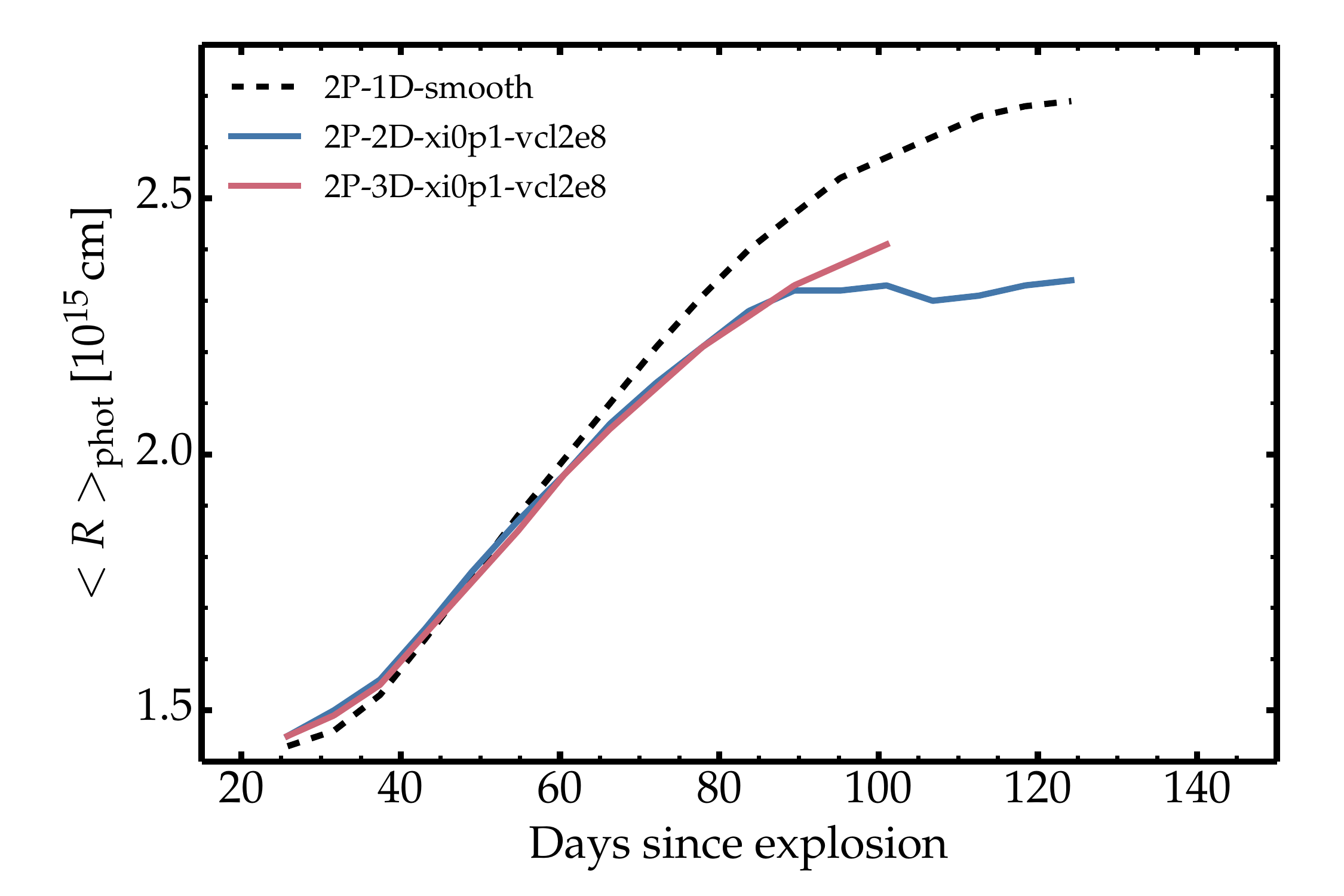}
   \end{center}
   \caption{Evolution of the angle-averaged photospheric radius for the 1D, 2D, and 3D simulations shown in Fig.~\ref{fig_lbol_2P}.
   \label{fig_rphot}
   }
   \end{figure}

   \begin{figure*}[h]
   \begin{center}
\includegraphics[width=0.495\hsize]{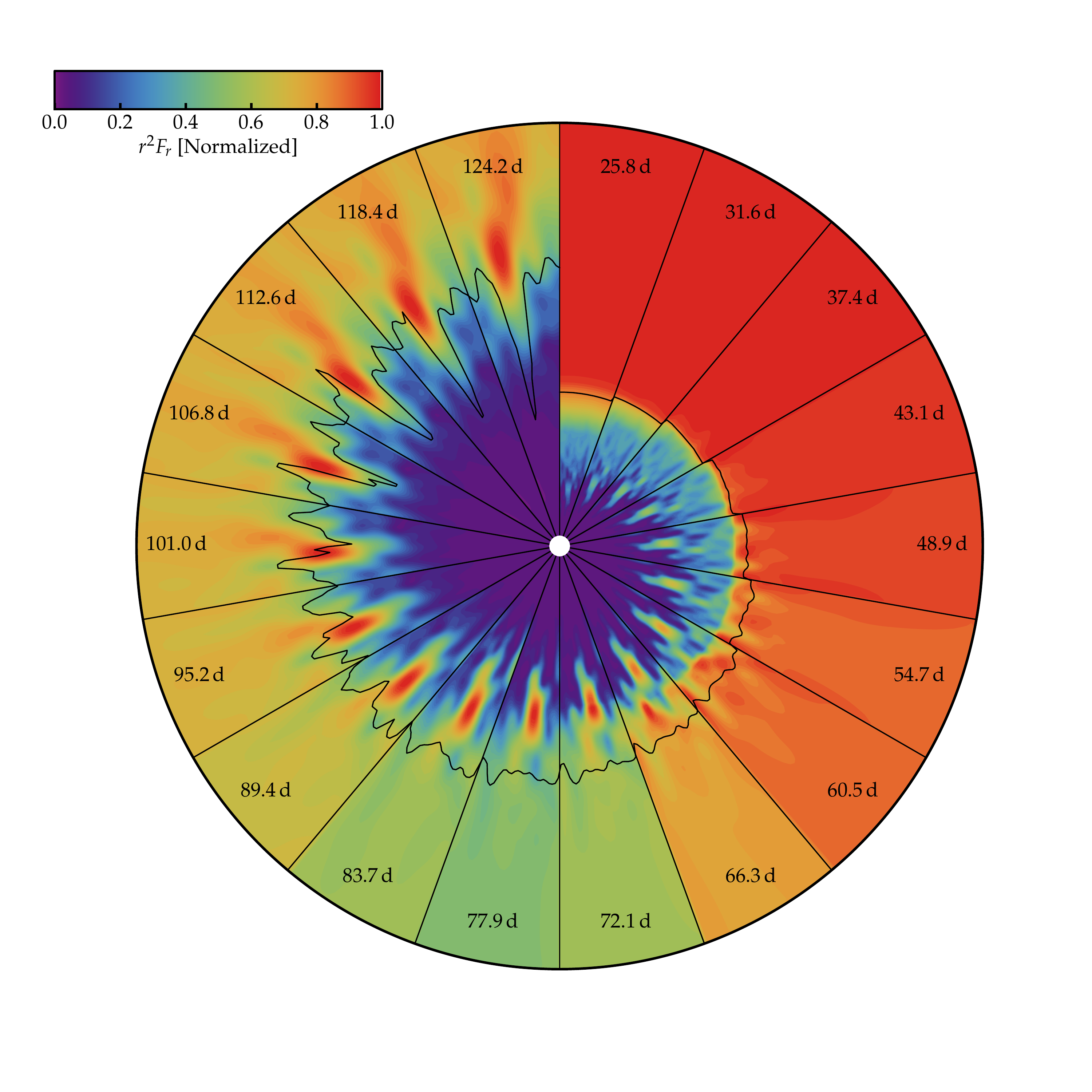}
\includegraphics[width=0.495\hsize]{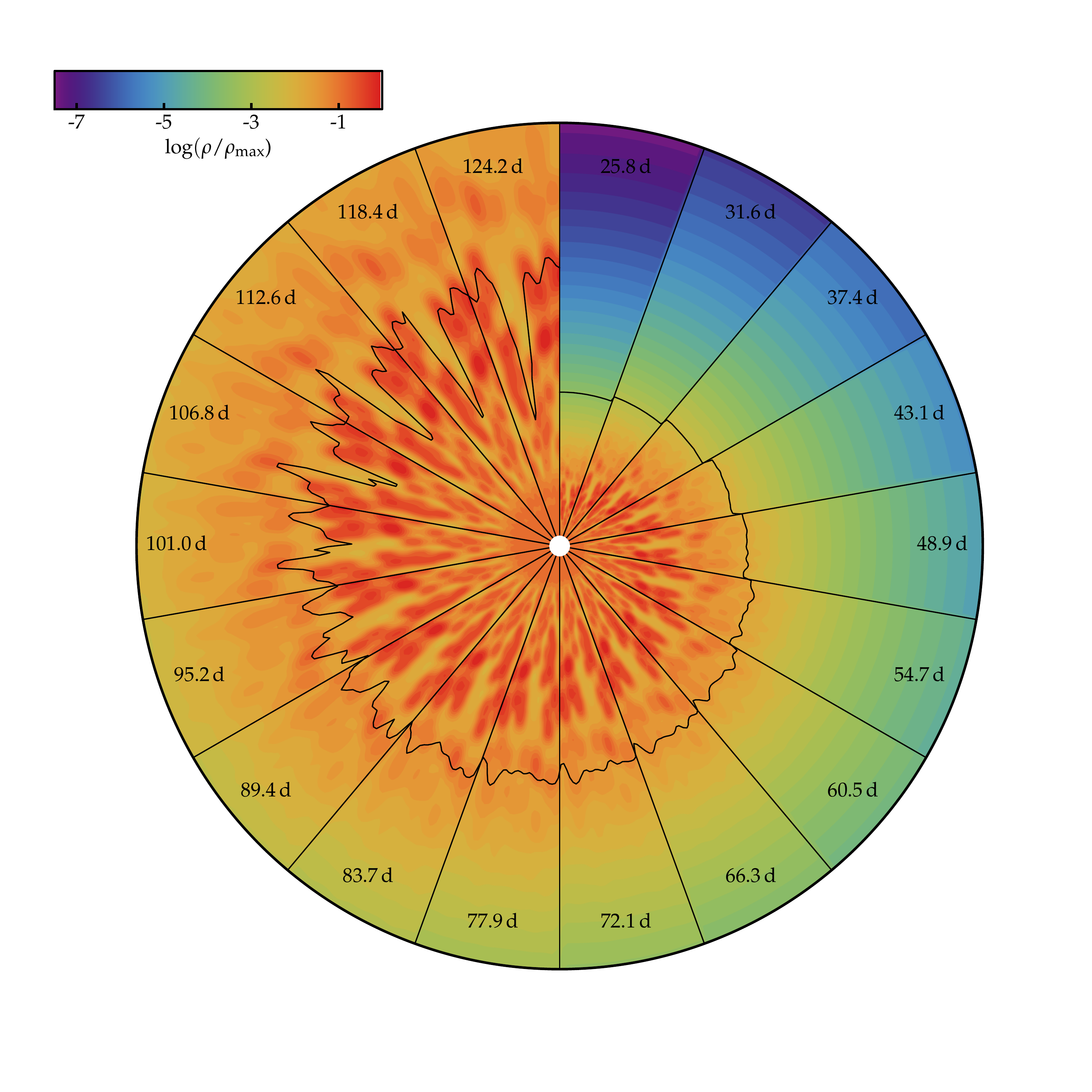}
\end{center}
   \caption{Clock plots for the radial radiative flux (left; the quantity shown is $R^2 F_r$, normalized to its maximum value at each time) and of the mass density $\rho$ (right; also normalized) as a function of time for the 2D simulation 2P-2D-xi0p1-vcl2e8 shown in Figs.~\ref{fig_lbol_2P}--\ref{fig_rphot}. Each wedge corresponds to one post-explosion epoch, starting at noon and progressing clockwise, with a time increment of 5.78\,d. In each panel, the black line corresponds to the photosphere (the maximum grid radius is $4 \times 10^{15}$\,cm).
 \label{fig_clock_plots}
}
\end{figure*}

 \section{Results for a reference case}
 \label{sect_ref}

Figure~\ref{fig_lbol_2P} shows the bolometric light curve for the 2P model simulated in 1D (smooth ejecta density structure) or in 2D and 3D (clumped structure). The adopted clumping is strong at low velocity but rises quickly to unity beyond a few thousand \kms. The clumping parameters are $\xi_0=$\,0.1, $V_{\rm cl}=$\,2000\,\kms, $\Delta V_{\rm cl}=$\,2000\,\kms, and $\sigma_{\rm cl}=$\,0.02.

Because this clumping does not affect the outermost ejecta layers, and because all models have the same \nifs\ mass (i.e., irrespective of clumping), the model luminosity is the same at the start of the simulation (when the photosphere is in the smooth outer ejecta layers) and at nebular times (when the total luminosity equals the total decay power, i.e., $L_{\rm bol} = L_{\rm decay}$). However, because of the different ejecta density structures (i.e., smooth or clumped), the rate at which the radiation is released from the ejecta differs between models.

In the clumped models, the early-time luminosity is greater than in the 1D smooth model for up to about 65\,d, after which it is below the predictions for the 1D model. The clumped models also transition earlier to the nebular phase. In our setup, the original ejecta temperature varies with radius but is independent of angle. The initial temperature structure is independent of clumping, and so does the total radiative energy stored in the ejecta. This energy is of the form $\int a_R T^4 dv$, where $a_R$ is the radiation constant and $dv$ is a volume element. The different models in Fig.~\ref{fig_lbol_2P} therefore radiate roughly the same time-integrated luminosity (modulated by expansion losses), but clumping influences the rate at which the stored radiative energy is released. Because the radiative energy is stored uniformly in volume, the segregation of mass into clumps lowers the trapping efficiency of the ejecta. The radiative flux is boosted between the clumps and the stored radiative energy can escape more freely.

Figure~\ref{fig_fr_rho} illustrates this effect for three comoving velocities of 2000, 3000, and 4000\,\kms\ for model 2P-2D-xi0p1-vcl2e8. Where the optical depth is large (i.e., at smaller velocities and earlier times), the normalized flux is impacted by the change in photon mean free path caused by clumping. Although the flux is not large at high optical depth, the flux contrast between clump and interclump medium is large. As we go to lower optical depth, the contrast between clump and interclump medium is unchanged at a given velocity, but the lateral fluctuations in radiative flux is weaker. The impact of clumping is most pronounced in the vicinity of the photosphere. Beyond the photosphere, the photon mean free path is larger so the material cannot trap efficiently the radiation, whether it is clumped or not. Below the photosphere, at high optical depth where the photon mean free path is small, the modulations caused by clumping have a weak influence on radiation leakage. But at moderate optical depth, the presence of clumping can allow radiation to leak out from between the clumps when the clumps are still optically thick. In this region, clumping can foster an earlier escape of radiation.

Consequently,  the angle-averaged photospheric radius increases more slowly (i.e., the photosphere recedes faster in mass space) in the 2D and 3D clumped models compared to the smooth 1D counterpart (Fig.~\ref{fig_rphot}). The greater recession is what causes the boost to the bolometric light curve. The effect is analogous to that caused by microclumping \citep{d18_fcl} but the process is different. With the microclumping treated in 1D nonLTE radiative transfer, the recombination rate at the photosphere is enhanced, which lowers the ionization and helps the photosphere to recede in mass space. With the macroclumping treated in (LTE) multi-D radiative transfer, it is the enhanced radiative losses that increase the photospheric cooling and causes the faster photospheric recession. In reality, both micro and macroclumping should be present. Because their effect acts in the same sense, the combination of both forms of clumping should yield a greater influence on the light curve than when only one form of clumping is present.

The evolution of the radiative flux and mass density at multiple epochs, from the start of the 2D simulation until the nebular phase, is shown in Fig.~\ref{fig_clock_plots}. At the earliest epochs, the photosphere (which is sensitive to the downstream density structure) is essentially spherical. The impact of clumping is first born in optically thick regions. By $\sim$\,60\,d, clumping is present below, at, and beyond the photosphere but its influence has been felt since the start. At this epoch, the bolometric luminosity in the 2D and 3D simulations drops below the value in the 1D smooth model (Fig.~\ref{fig_lbol_2P}). These ``clock plots'' also show how radiation progresses more efficiently through the lower density regions between the clumps (see also Fig.~\ref{fig_fr_rho}). The process is time dependent because of ejecta expansion and the depth-dependence of clumping, and also because the radiant energy is typically more abundant in the outer ejecta than in the inner ejecta. Different clumping properties, combined with different ejecta properties, would yield different behaviors.

Figure~\ref{fig_lbol_2P} also includes the bolometric light curve for the 3D model. Interestingly, going from 1D to 2D leads to a greater change to the light curve than going from 1D to 3D, even if the effect is qualitatively the same. The different quantitative behavior may result from the greater porosity of the ejecta in 2D since clumps are structured as tori, mimicking the effect of aligned clumps in 3D. When clumps are randomly distributed in 3D, they more efficiently cover ejecta-centered sight lines and thus better trap the stored radiation. These properties depend on the adopted clumping properties.

\section{Influence of some variations in clumping characteristics}
\label{sect_clump_var}

\begin{figure*}
     \includegraphics[width=0.49\hsize]{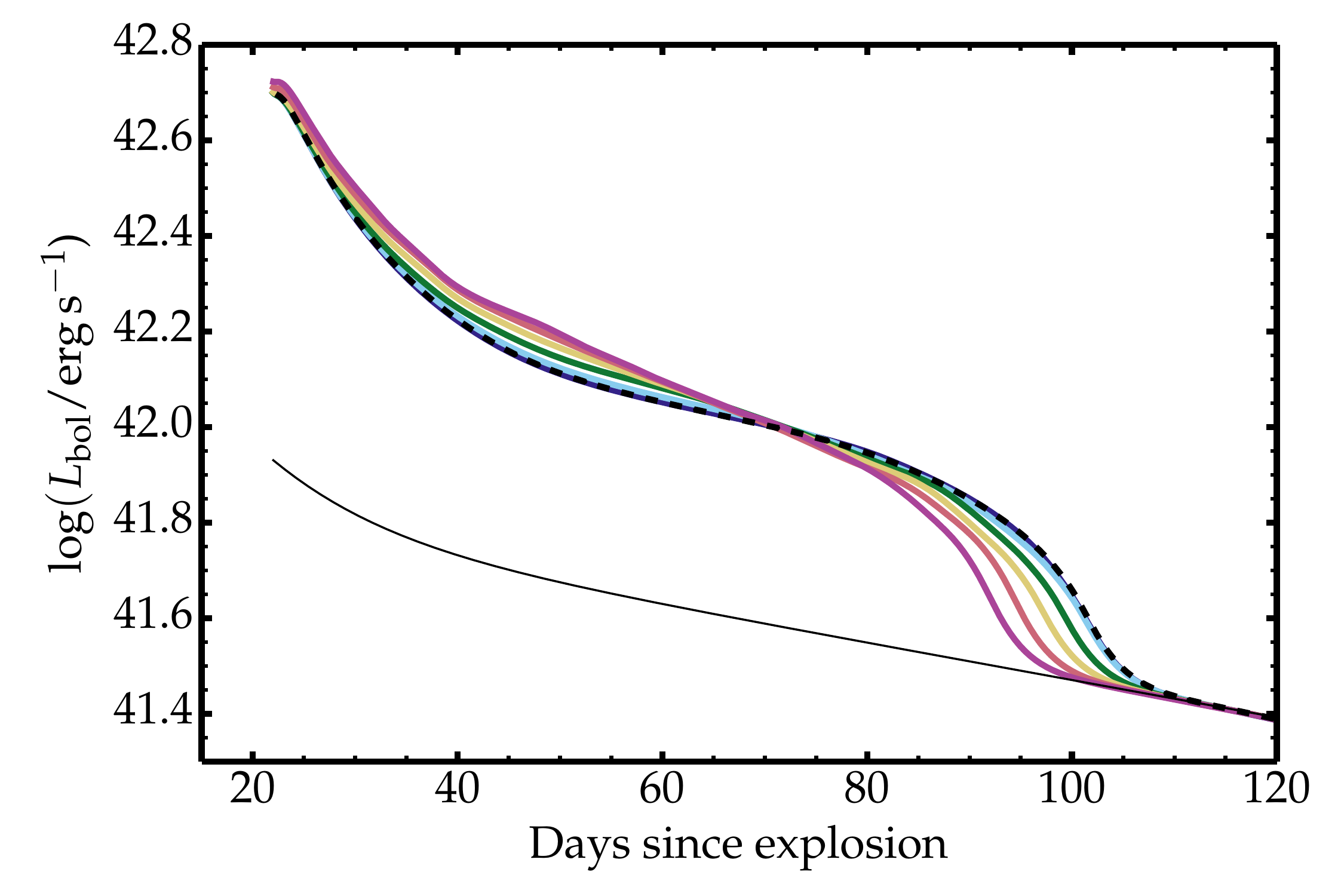}
     \includegraphics[width=0.49\hsize]{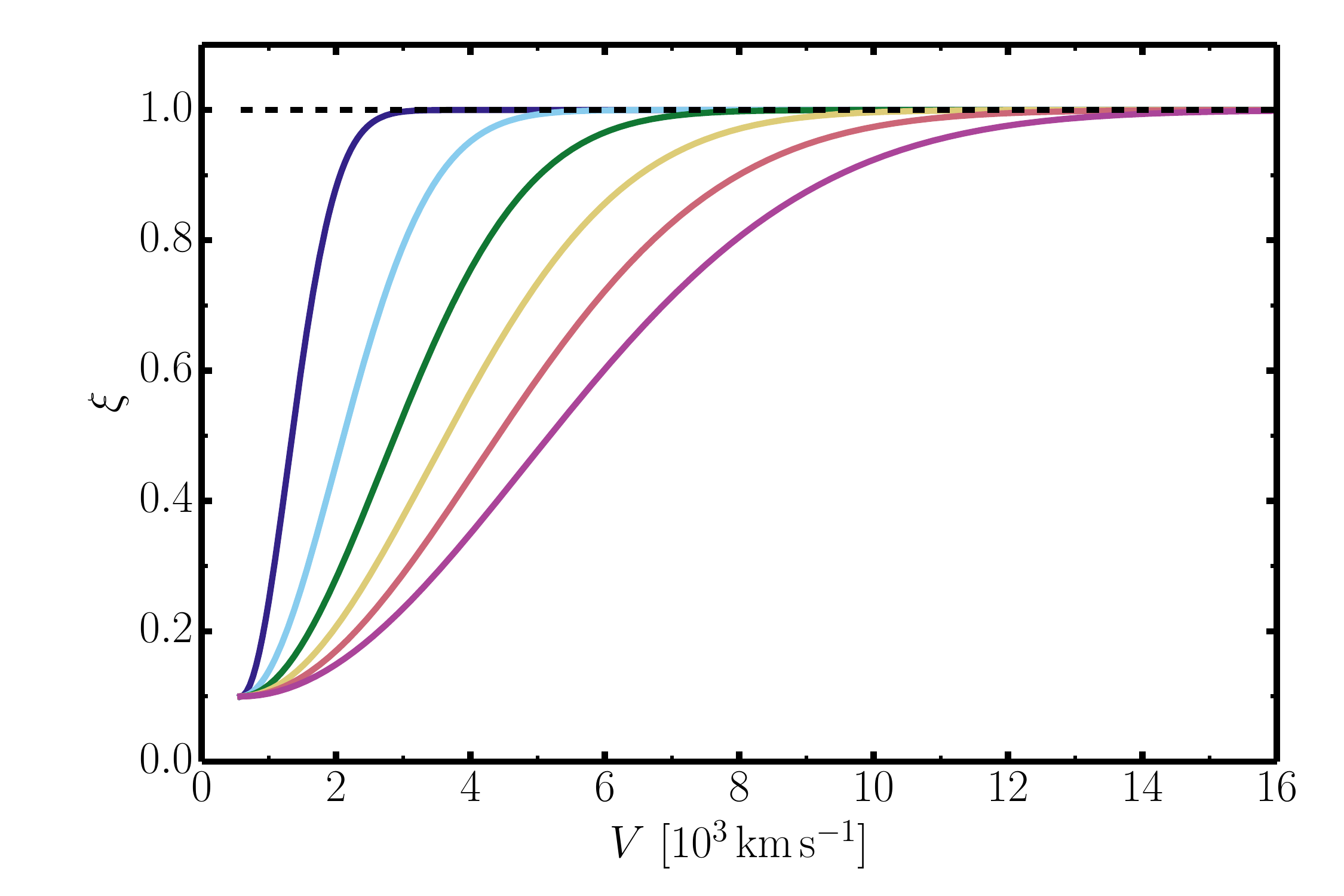}
   \caption{Bolometric light curves (left) for a set of 2D clumped models for the 2P case in which the radial clumping profile $\xi(V)$ is modified. The thin black line corresponds to the instantaneous decay power. For the clumping parameters, the present set uses $\xi_0=$\,0.1, $V_{\rm cl}=$\,0\,\kms, and $\Delta V_{\rm cl}$ ranges between 1000 and 6000\,\kms\ (right). The dashed line corresponds to the 1D smooth model counterpart.
\label{fig_vcl0}
   }
   \end{figure*}

\begin{figure*}
     \includegraphics[width=0.49\hsize]{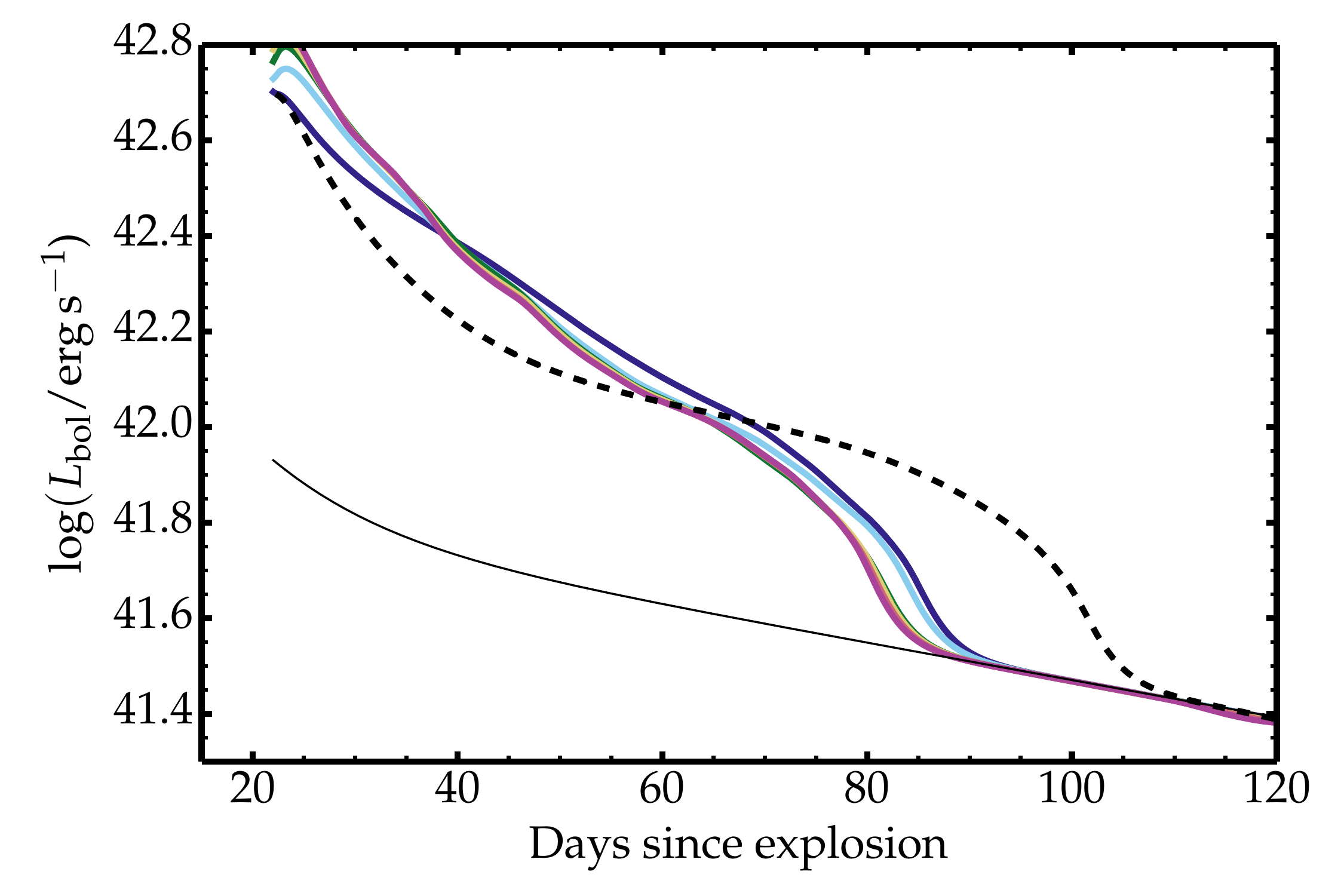}
     \includegraphics[width=0.49\hsize]{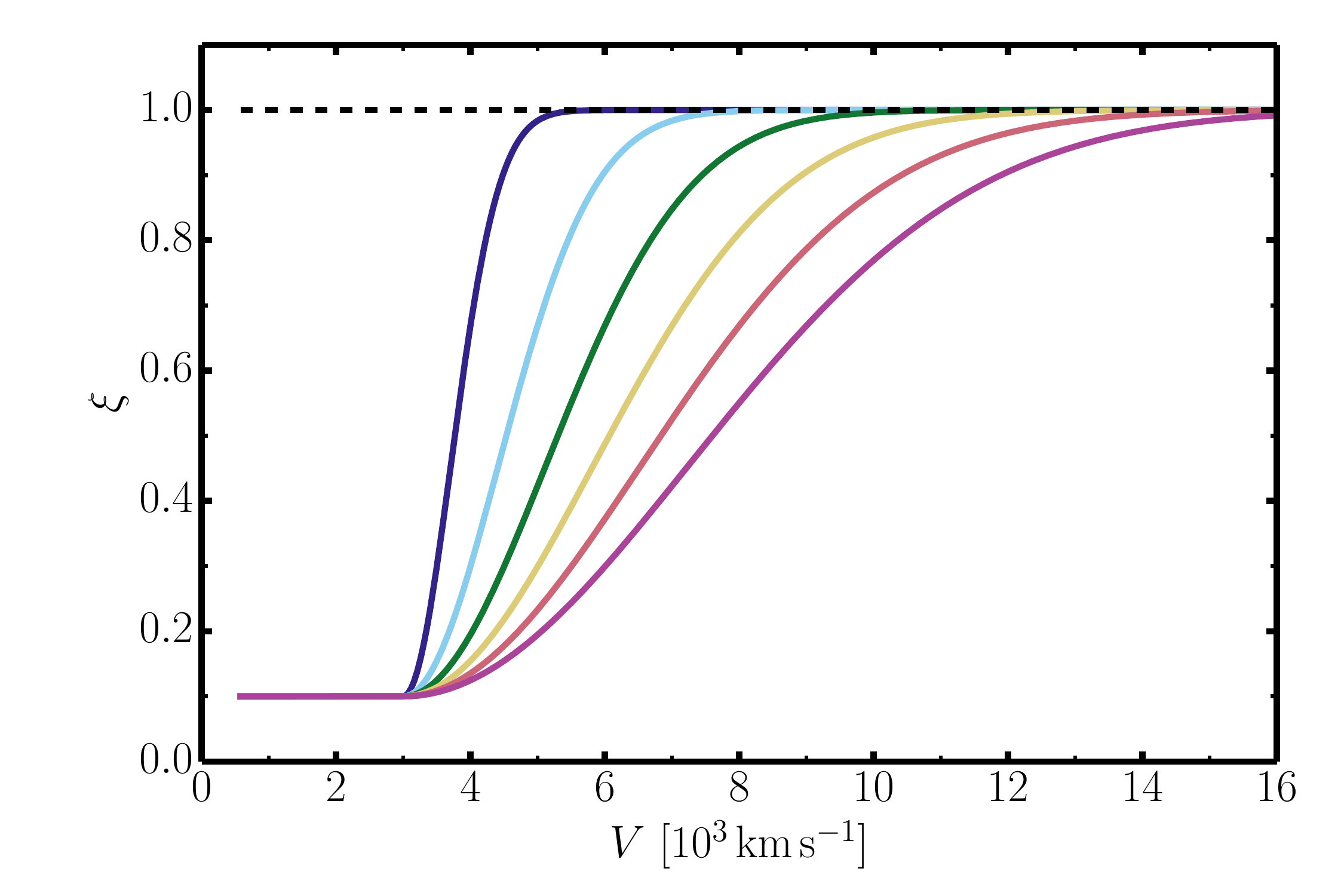}
   \caption{Same as Fig.~\ref{fig_vcl0}, but now using $V_{\rm cl}=$\,3000\,\kms
   \label{fig_vcl1}}
   \end{figure*}

Figures~\ref{fig_vcl0} and \ref{fig_vcl1} illustrate the impact of clumping properties on the resulting bolometric light curves of clumped 2D ejecta for the 2P ejecta conditions (Table~\ref{tab_grid}). For the models in Fig.~\ref{fig_vcl0}, the adopted clumping properties are $\xi_0=$\,0.1, $V_{\rm cl}=$\,0\,\kms, and $\Delta V_{\rm cl}$ ranges between 1000 and 6000\,\kms. In Fig.~\ref{fig_vcl1}, the ejecta properties are analogous except that $V_{\rm cl}=$\,3000\,\kms.

For the model with $V_{\rm cl}=$\,0\,\kms\ and $\Delta V_{\rm cl}=$\,1000\,\kms\ (Fig.~\ref{fig_vcl0}), the light curve is unaffected by clumping (it overlaps with that for the 1D smooth model). This arises because clumping is confined to the innermost ejecta layers, which contain very little radiative energy at $\sim$\,100\,d. Whether this material is clumped or not makes no difference since there is no energy to release. As $\Delta V_{\rm cl}$ is enhanced, clumping covers a larger range of the ejecta so a greater fraction of the volume that stores the radiative energy reacts to the change in photon mean-free path. A greater impact on the light curve is obtained when $V_{\rm cl}$ is raised from zero to 3000\,\kms. This velocity threshold corresponds roughly to the edge of the progenitor core (see Fig.~\ref{fig_init}).

In type II SN progenitors, clumping should be stronger in the inner ejecta, which corresponds to the shocked progenitor He core. This material, which contains less radiative energy than the shocked H-rich envelope, should thus be made even more transparent because of clumping, compromising even more the inference of its mass from light curve modeling. Clumping has a visible effect on the light curve only if it takes place within the H-rich layers of the progenitor where the bulk of the radiative energy is stored.

\begin{figure}
\includegraphics[width=\hsize]{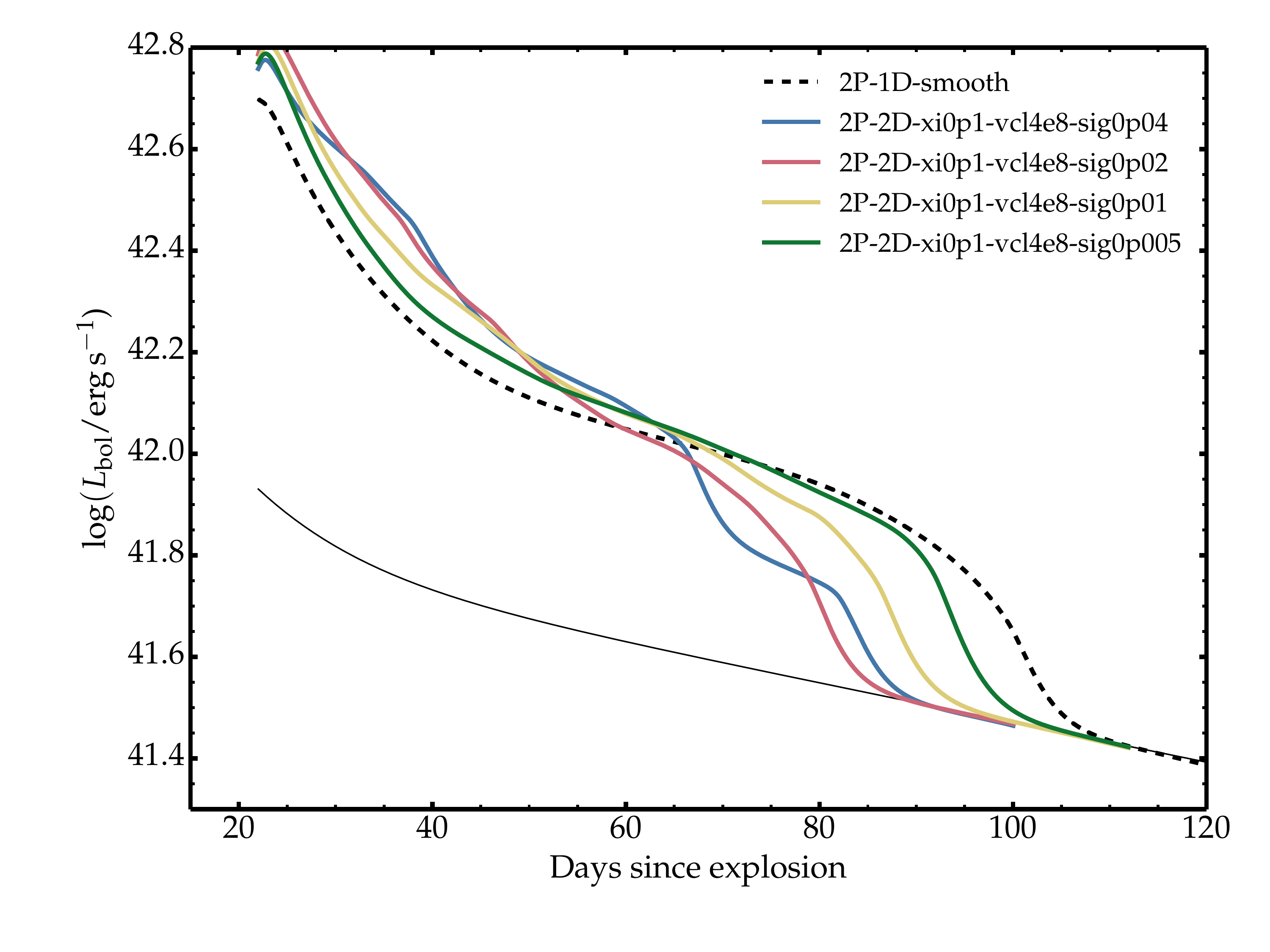}
\caption{Comparison of bolometric light curves for 2D 2P ejecta models with $\xi_0=$\,0.1,
$V_{\rm cl}=$\,4000\,\kms, and with different choices for the clump lateral size $\sigma_{\rm cl}$, which covers from 0.005 to 0.04 times the local radius (see label as well as Eq.~\ref{eq_rho_cl} for details).
\label{fig_clump_size}
}
\end{figure}

Figure~\ref{fig_clump_size} shows the impact of the adopted clump size on the bolometric light curve. The larger are the clumps, the stronger is the impact on the light curve, with a great boost at early times, a shorter optically-thick phase duration (earlier transition to the nebular phase), and greater fluctuations.  What drives this behaviour is that the number of clumps drops as their size is increased, facilitating the escape of radiation between fewer and larger clumps. For smaller clumps, the lines of sight are more evenly covered by clumps, preventing the escape of radiation.  For an infinitely small clump size, the radial optical depth converges towards that for the smooth case and the trapping efficiency of the material is unchanged. This is the case described in \citet{d18_fcl}, in which clumping acts primarily through its influence on the recombination rate.

\section{Influence of chemical inhomogeneities}
\label{sect_he}

We have also used our clumping-formalism to mimic chemical segregation. Isolated regions of space (i.e. clumps) were turned into pure helium, while the surrounding material was a mixture of H and He. In this simulation of chemical segregation, the density was untouched and thus identical to the 1D (smooth) model counterpart. Within the simulation, this compositional difference was conveyed through a distinct recombination temperature, set to 10,000\,K for the He-rich material and 6000\,K for the rest -- the actual material opacity was kept the same as for the solar metallicity mixture because the reduced electron-scattering contribution in a He-rich plasma is compensated by the greater contribution from metal lines. The critical feature of He-rich (H-poor) material is that it recombines at higher temperatures than H-rich material.

For simplicity, we used the 2P ejecta model with no \nifs\ and thus no decay power. We performed one simulation in 2D, no clumping, and uniform composition (the model name is 2P-2D-smooth). A second simulation was done with the same setup, but in which 30 blobs of pure helium were randomly distributed in both latitude and radius, between the innermost ejecta layer and 5000\,\kms.
The spatial extent of the He blobs is 2\% of the local radius (the blobs are tori in 2D).

\begin{figure}
     \includegraphics[width=\hsize]{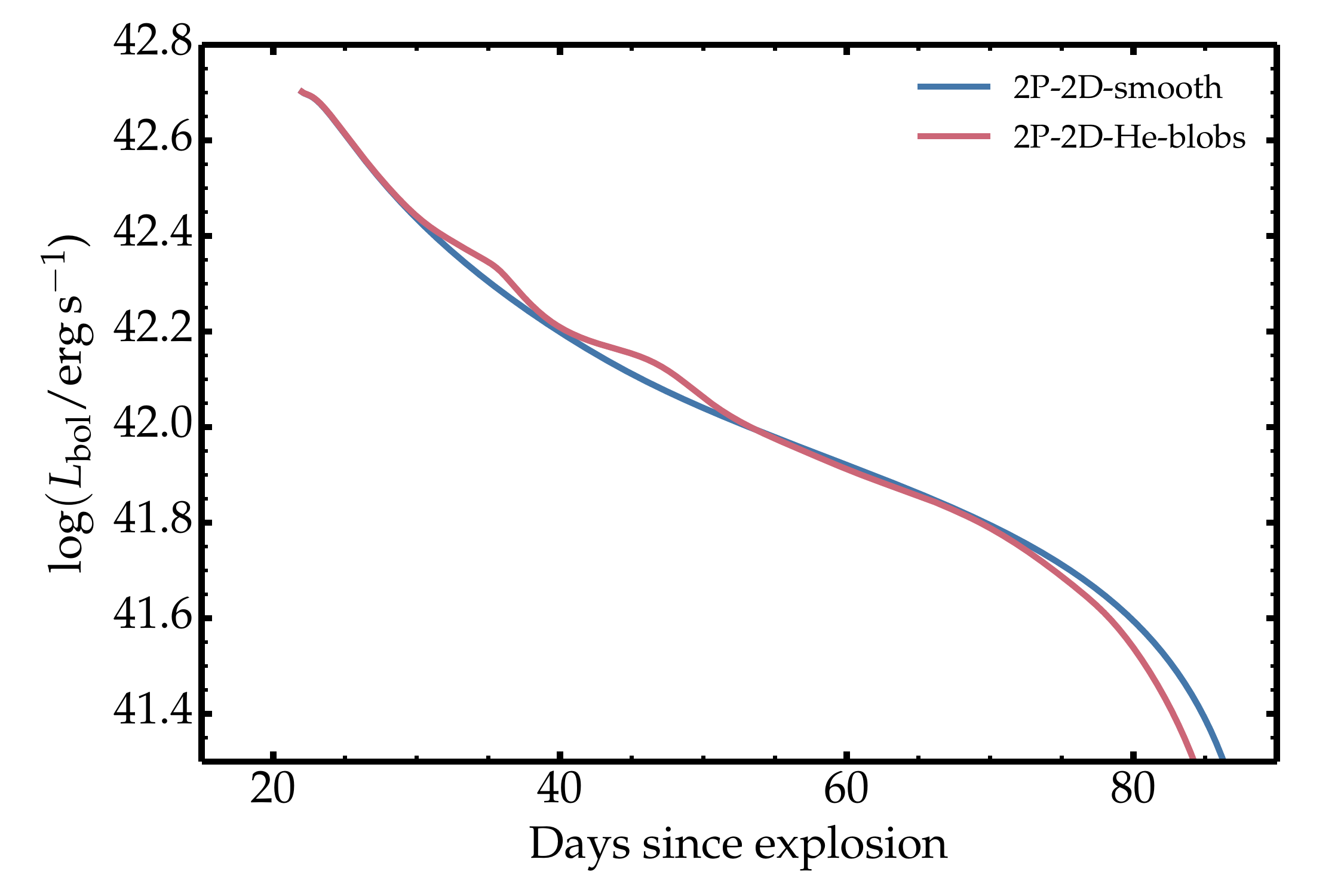}
   \caption{Bolometric light curves for two 2P ejecta models simulated in 2D, one being smooth and of uniform (solar-metallicity) composition, the other including 30 pure-He blobs (with a characteristic size set to $0.02\,R$) randomly distributed up to a velocity of 5000\,\kms. Radioactive decay power is ignored in this set of simulations.
      \label{fig_he_blob}}
   \end{figure}

In the 2D \heracles\ simulation with such He-rich blobs, the light curve presents two broad bumps and a slightly faster transition to the nebular phase compared to the 2D smooth (homogeneous) ejecta (Fig.~\ref{fig_he_blob}). This feature is caused by the faster recombination in the He-rich blobs, hastening the recession of the photosphere and the release of stored energy (from within and below the blobs).  Since the blobs do not change the radiative energy budget of the ejecta, the slightly greater release of energy early on leads to a faster transition to the nebular phase. The effect found here in 2D with He-rich blobs is similar to the results of 1D simulations by \citet{khatami_kasen_lc_19} in which the recombination temperature of the ejecta material is increased.

In Nature, type II SN light curves could present fluctuations associated with the presence of chemical inhomogeneities. This depends also on the size, number, and composition of such inhomogeneities. Their presence is very likely a result from the chemical mixing caused by Rayleigh-Taylor instabilities and post-shock neutrino-driven convection (see, e.g., \citealt{wongwathanarat_13_3d}).

\section{Results for a more compact, BSG, progenitor}
\label{sect_bsg}

     We have also explored the influence of macroclumping in an ejecta produced by the explosion of a more compact progenitor, namely a BSG star. The initial conditions (especially for the temperature) for the calculation were adjusted to deliver a bolometric light curve similar to that of SN\,1987A \citep{catchpole_87A_87,hamuy_87A_88} -- see also Section~\ref{sect_grid_pres} and Table~\ref{tab_grid}.

     For the 2pec set, the simulations are started a little earlier, at 11\,d, when the ejecta has already started to recombine, but not too early so that it has expanded sizably. To resolve this more compact structure initially, we used a higher resolution, with 960 radial zones, and 96 zones in $\theta$  --- no 3D simulation was performed because there are too costly for our computer capabilities.

\begin{figure}
   \begin{center}
\includegraphics[width=\hsize]{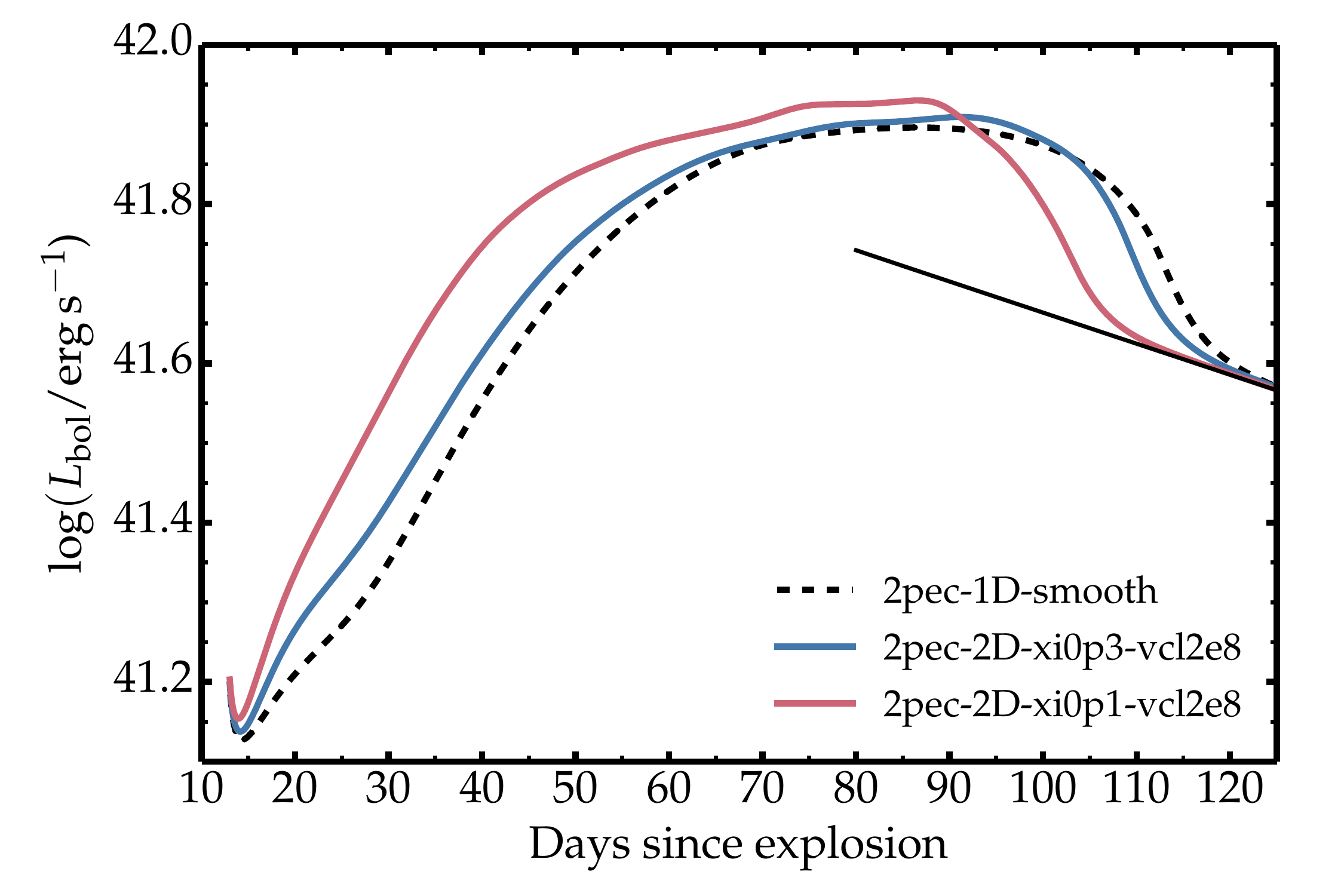}
\end{center}
   \caption{Same as Fig.~\ref{fig_lbol_2P}, but now for the 2pec models. For the clumped models, the parameters are $V_{\rm cl}=$\,2000\,\kms\ and $\Delta V_{\rm cl}=$\,2000\,\kms, with $\xi_0=$\,0.3 (0.1) in model 2pec-2D-xi0p3-vcl2e8 (2pec-2D-xi0p1-vcl2e8). The solid black line gives the instantaneous decay rate from an initial \nifs\ mass of 0.078\,\msun.
\label{fig_2pec}
   }
   \end{figure}

     Figure~\ref{fig_2pec} shows the bolometric light curve for two clumped models and the 1D-smooth
counterpart, all using the 2pec ejecta parameters shown in Table~\ref{tab_grid}. For the clumped models, the parameters are $V_{\rm cl}=$\,2000\,\kms\ and $\Delta V_{\rm cl}=$\,2000\,\kms, with $\xi_0=$\,0.3 (0.1) in model 2pec-2D-xi0p3-vcl2e8 (2pec-2D-xi0p1-vcl2e8). As expected, with clumping, the bolometric luminosity rises faster because radiation leakage is facilitated. The greater luminosity early on corresponds to a greater recession of the photosphere, which leads to a shorter optically-thick phase. The model with the greater clumping turns nebular $10-15$\,d before the 1D smooth counterpart. The effect of clumping here is analogous to the effect of \nifs\ mixing in BSG explosion models (see, e.g., \citealt{blinnikov_87A_00}), although for different reasons. With \nifs\ mixing, power is generated further out in the ejecta, circumventing the long delay otherwise needed for diffusion (or for the recession of the photosphere into the layers rich in \nifs). In contrast, with clumping, the recession of the photosphere and the diffusion of stored radiation are both hastened.

   \begin{figure*}
   \begin{center}
\includegraphics[width=\hsize]{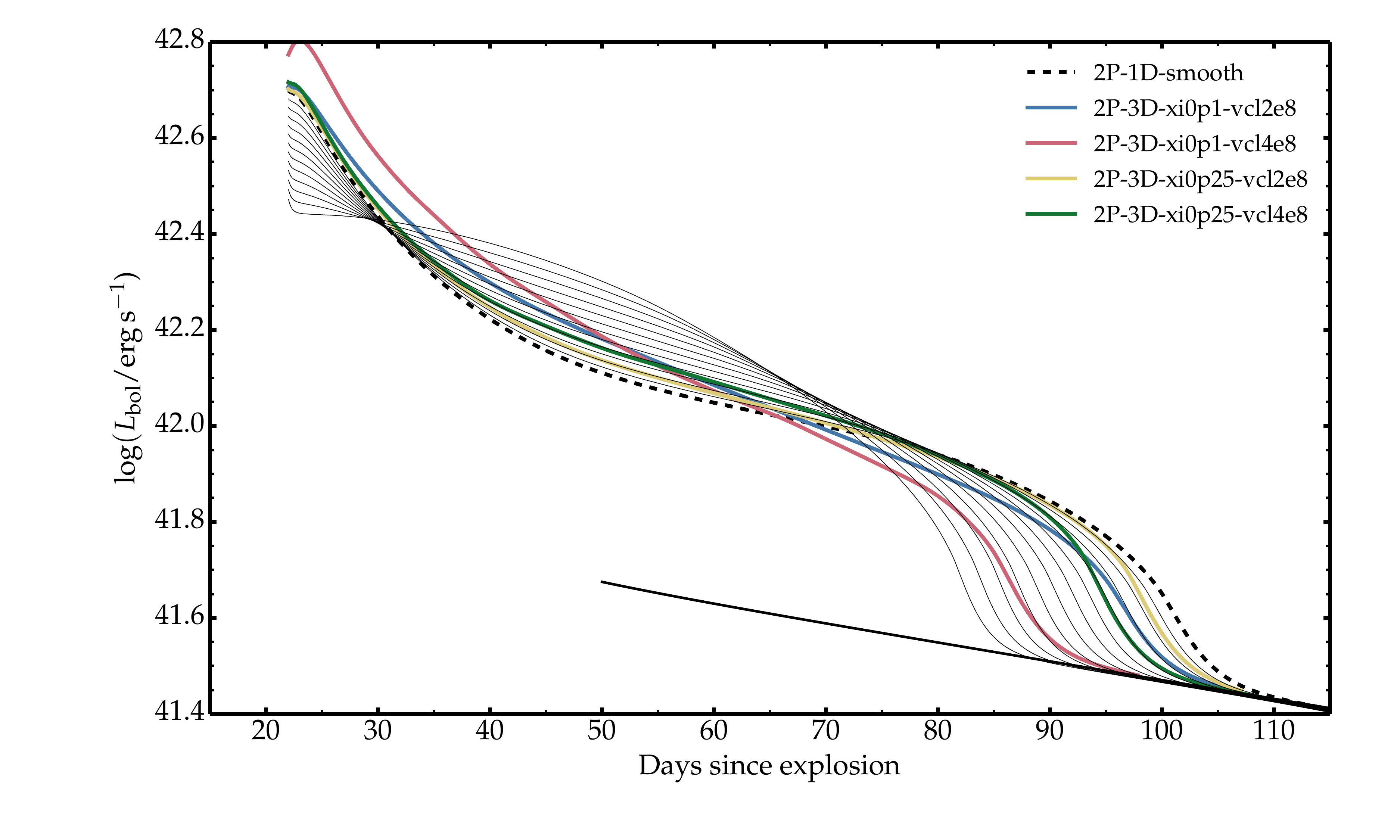}
\end{center}
\vspace{-1cm}
   \caption{Same as Fig.~\ref{fig_lbol_2P}, but now comparing the results for the 12\,\msun\ ejecta model 2P-1D-smooth (thick dashed) with those from the 3D counterparts in which the clumping parameters $\xi_0$ and $V_{\rm cl}$ are varied (thick colored lines). We overlay the results from additional 1D smooth models in which the ejecta mass is progressively decreased by steps of 0.4\,\msun\ from 12\,\msun\ down to 7.6\,\msun\ (thin black lines), which produces a continuous sequence of events with a shorter photospheric-phase duration than model 2P-1D-smooth. [See Section~\ref{sect_mej} for discussion.]
   \label{fig_1D_3D}
   }
   \end{figure*}

\section{Implications for the inferred type II ejecta masses}
\label{sect_mej}

All the simulations presented in the preceding sections show that clumping leads to a shortening of the photospheric phase. A clumped ejecta may thus appear as a smooth ejecta of a lower mass. In this section, we try to quantify this effect by running a restricted set of simulations for 3D clumped ejecta and compare the resulting light curve to smooth (1D) ejecta having a lower mass.

To limit the parameter space, we first performed a set of 2D clumped simulations based on the 2P ejecta model parameters (Table~\ref{tab_grid}) in which we varied the parameters $\xi_0$ (values 0.1, 0.25, and 0.5)  and $V_{\rm cl}$ (values 1000, 2000, 3000, and 4000\,\kms). All combinations of $\xi_0$ and $V_{\rm cl}$ were done. The results showed that models with $\xi_0=$\,0.5 or $V_{\rm cl}=$\,1000\,\kms\ are very similar to the 1D smooth result. Similarly, the results for $V_{\rm cl}$ of 3000 and 4000\,\kms\ are very close to each other. Hence, discarding these superfluous choices, we performed 3D simulations for four cases only, using $\xi_0$ of 0.1 and 0.25, and $V_{\rm cl}$ of 2000 and 4000\,\kms\ (the 3D model with $\xi_0=$\,0.1 and $V_{\rm cl}=$\,2000\,\kms\ was already presented in Section~\ref{sect_ref}).

As discussed in the preceding sections and now demonstrated in the 3D clumped simulations shown in Fig.~\ref{fig_1D_3D}, decreasing $\xi_0$ or increasing $V_{\rm cl}$ leads to an enhancement of the early-time luminosity and a shortening of the photospheric phase. Figure~\ref{fig_1D_3D} also shows the light curves for 1D smooth model counterparts in which the ejecta mass is reduced. The morphology of the resulting set of light curves differs from that of the 12\,\msun\ ejecta simulations because the association between density and temperature (i.e., for a given profile $T(R)$) is no longer the same. However, the overall behavior is similar in the sense that the lower ejecta masses yields larger luminosities early on (i.e. during the first two third of the photospheric phase, but with the exception of the first week) and an earlier transition to the nebular phase since the ejecta contains the same amount of trapped radiation energy initially.

If we use the duration of the photospheric phase as a proxy for ejecta mass,  our 3D clumped models correspond to various reductions in ejecta mass compared to a 1D smooth ejecta model with identical parameters (as given Table~\ref{tab_grid}). Model 2P-3D-xi0p25-vcl2e8 matches the 1D smooth model that is 0.5\,\msun\ less massive. For models 2P-3D-xi0p25-vcl4e8 and 2P-3D-xi0p1-vcl2e8, this reduction is $1-1.5$\,\msun. For model 2P-3D-xi0p1-vcl4e8, the reduction is about 3.5\,\msun.

This exploration is somewhat artificial since one cannot in principle change the ejecta mass without changing the other quantities. But it suggests that clumping, if strong and extended, can lead to a significant underestimate of the ejecta mass. The magnitude of the effect on the light curve depends on the level of clumping in the ejecta regions that contain a sizable amount of trapped radiation. Hence, it depends both on the density structure (average density and clumping profile) and on the temperature structure (how the radiation energy is distributed in velocity or mass space).

\section{Conclusions}
\label{sect_conc}

We have presented a set of gray radiation-hydrodynamics simulations in 1D, 2D, and 3D for type II SN  ejecta from BSG and RSG progenitors. The simulations are limited to the phase of homologous expansion, starting at $10-20$\,d after explosion. For simplicity, the initial ejecta conditions are set analytically using guidance from more sophisticated radiative transfer simulations of type II SNe. Macroclumping is introduced in the form of radially and laterally confined high-density regions (tori in 2D, spheroids in 3D), with an extent set to some fraction (typically $\sim$\,2\%) of the local radius.

Macroscopic clumping acts to boost the radiation leakage from type II SN ejecta by enhancing the escape of radiation between overdense regions. The effect can be strong because at a given radius or velocity, the clumps and the interclump medium have the same temperature in optically-thick regions, given by what was produced by the shock and subsequently degraded by expansion. By segregating more and more mass into dense clumps, a greater amount of stored radiation becomes trapped within a lower density medium (the interclump medium), so that its escape is facilitated. The general effect is thus to boost the early time luminosity and shorten the photospheric phase. In BSG explosions, macroscopic clumping also leads to a shorter rise time to bolometric maximum.

The exact impact of clumping on the SN radiation depends on numerous aspects. The effect of clumping increases as the size of the clumps increases, which also tends to introduce large amplitude fluctuations in the light curve. For small enough clumps, the medium acts as if it was smooth (if we neglect the influence on the recombination rate and the ionization; see \citealt{d18_fcl}). Although potentially stronger, clumping at low velocity has little impact on the light curve during the photospheric phase because the inner ejecta layers contain only a small fraction of the total radiation budget. The larger volume occupied by the ejecta regions at larger velocities stores more radiation energy, but clumping is expected to be weaker there.  Hence, it is the clumping at intermediate velocities of a few 1000\,\kms\ that probably has the strongest impact on type II SN light curves. In our set of simulations, the strongest impact was obtained for cases in which clumping corresponded do a maximum density contrast of a few tens out to about 4000\,\kms. In the case of strong clumping, the 3D clumped model showed a light curve analogous to that of a smooth ejecta model with a 30\% lower ejecta mass. For lower values of clumping, the offset in ejecta mass may be only 10 or 20\%.

Clumping may also appear in the form of composition inhomogeneities rather than density variations. An interesting effect is the case of He-rich clumps since their recombination temperature is larger than for a mixture of H and He. In 2D \heracles\ simulation with such He-rich blobs, the light curve presents low-frequency variations compared to the smooth (homogeneous) ejecta counterpart, as well as a slightly earlier transition to the nebular phase. In reality, there may be simultaneously density variations and chemical inhomogeneities, with distinct properties at different depths, yielding a complicated effect on the SN light curve.

Clumping may also be associated with a microscopic effect, not included in the present simulations, but discussed in \citet{d18_fcl}. With clumping, the recombination rate (which scales with the square of the local gas density) is boosted so that dense clumps will recombine on a shorter time scale than the surrounding lower density medium. In practice, a clumped medium will be a complicated mixture of regions with different density, temperature, ionization (as well as composition if chemical segregation is taken into account). But clumping should lead to a faster recession of the photosphere through the lower-density interclump medium as well as the lower-temperature lower-ionization higher-density clumps. Clumping is probably present on a variety of scales in SN ejecta, but on all scales, clumping tends to facilitate the release of stored energy. Quantitatively, the simulations in this study suggest that large-scale clumping may not significantly impact type II SN light curves  because this requires density contrasts of a few tens between clump and interclump medium. A greater impact on SN observables may arise from a microphysical effect of clumping, through a boost of recombination rates, which can occur for density contrasts of a few rather than a few tens \citep{d18_fcl}.

Observationally,  micro- and macroclumping may be at the origin of some of the diversity of type II SNe, including visual decline rates and photospheric phase duration (see e.g. \citealt{anderson_2pl}), colors (see e.g. \citealt{dejaeger_2p_col_18}), and spectral peculiarities (see e.g. \citealt{d19_sn2pec} for the case of Ba\two\ lines in SNe II-pec). Treating both micro- and macroclumping in a given type II SN simulation is challenging since it requires both nonLTE, time-dependence, and multi-D radiation transport, something that is not currently doable. Both clumps and interclump medium need to be explicitly modeled since these regions of different density (both at a given radius and at different depths) will have different temperatures and ionization levels (even for the same composition), hence different opacity and emissivity.

In the future, we will investigate the effect of clumping in Type Ibc SNe. These ejecta are distinct from type II SNe since the radiated energy arises more strongly from the continuous decay of unstable isotopes rather than from the release of initially stored shock-deposited energy. Clumping may nonetheless facilitate radiation escape and impact our inference of SN Ibc ejecta masses.

\begin{acknowledgements}
LD thanks ESO-Vitacura for their hospitality. This work was granted access to the HPC resources of  CINES under the allocation  2018 -- A0050410554 made by GENCI. We thank John Hillier for fruitful  discussions.
\end{acknowledgements}

\appendix

\section{Influence of the albedo}
\label{appendix_albedo}

\begin{figure}
   \begin{center}
\includegraphics[width=\hsize]{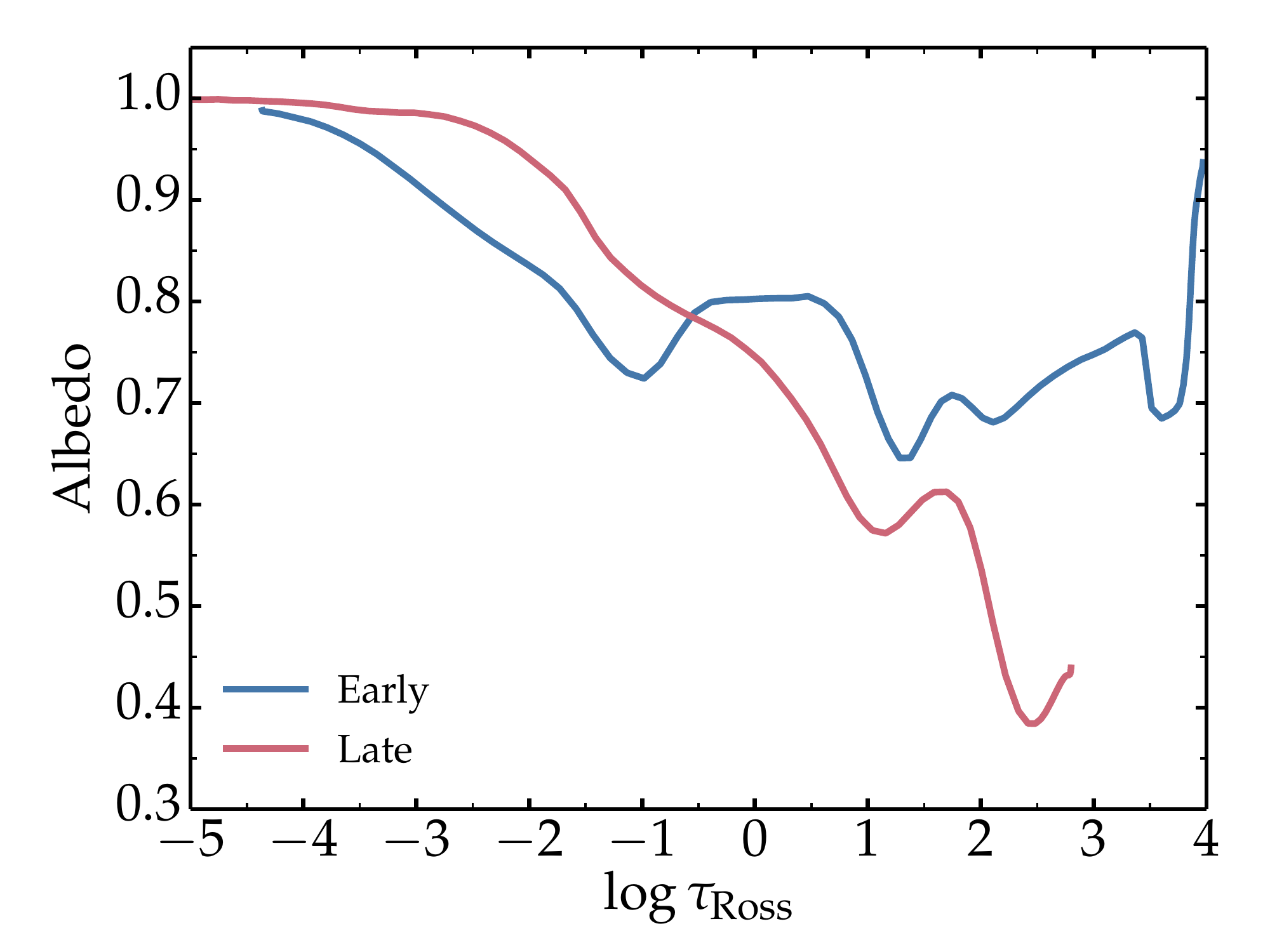}
\end{center}
   \caption{Variation of the albedo in a type II SN at early and late time during the photospheric phase. In this study, we use of value of 0.9, which is comparable to what holds in the photospheric regions where takes place the bulk of the radiative diffusion that influences the light curve.}
   \label{fig_chi_albedo}
\end{figure}

 In the present study, all simulations are performed using an albedo (defined as the ratio of scattering opacity to total opacity) of 0.9 throughout the ejecta. This is taken as a representative value for type II SN ejecta, and this approximation seems suitable given that we already make approximations in our modeling by assuming a grey opacity and neglecting nonLTE and time-dependent effects. Nonetheless, to support this choice we show in Fig.~\ref{fig_chi_albedo} how the albedo (we use here the ratio of the electron scattering opacity to the Rosseland-mean opacity) varies with Rosseland-mean optical depth at two epochs (early is before recombination, at 16\,d past explosion, and late is during the recombination phase, at 70\,d after explosion) in a Type II SN ejecta simulated with \cmfgen\ (see, for example, \citealt{d13_sn2p}). The conditions at these two epochs and for similar SN models bracket the range of ejecta conditions simulated here with \heracles. In the outer regions, the albedo is high due to ionization freeze out and the low density (the Rosseland mean opacity does not make much sense physically because conditions are nonLTE). At great depth, the albedo is low at late times because of the large contribution from line opacity, while at early times, the conditions are strongly ionized and electron-scattering dominates. In the photospheric regions, the albedo is around 0.8, which is quite close to our choice of 0.9.

\begin{figure*}
   \begin{center}
\includegraphics[width=0.49\hsize]{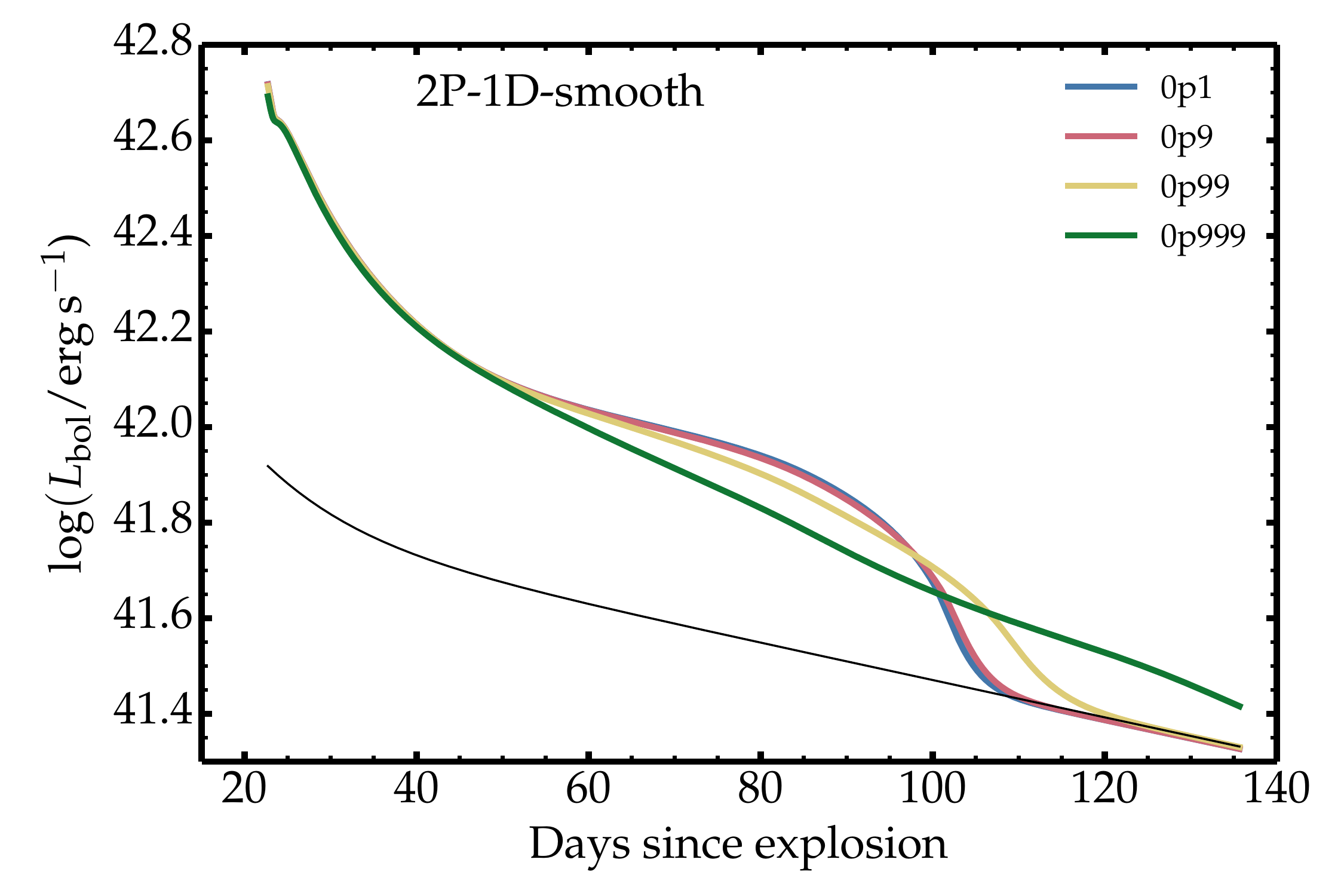}
\includegraphics[width=0.49\hsize]{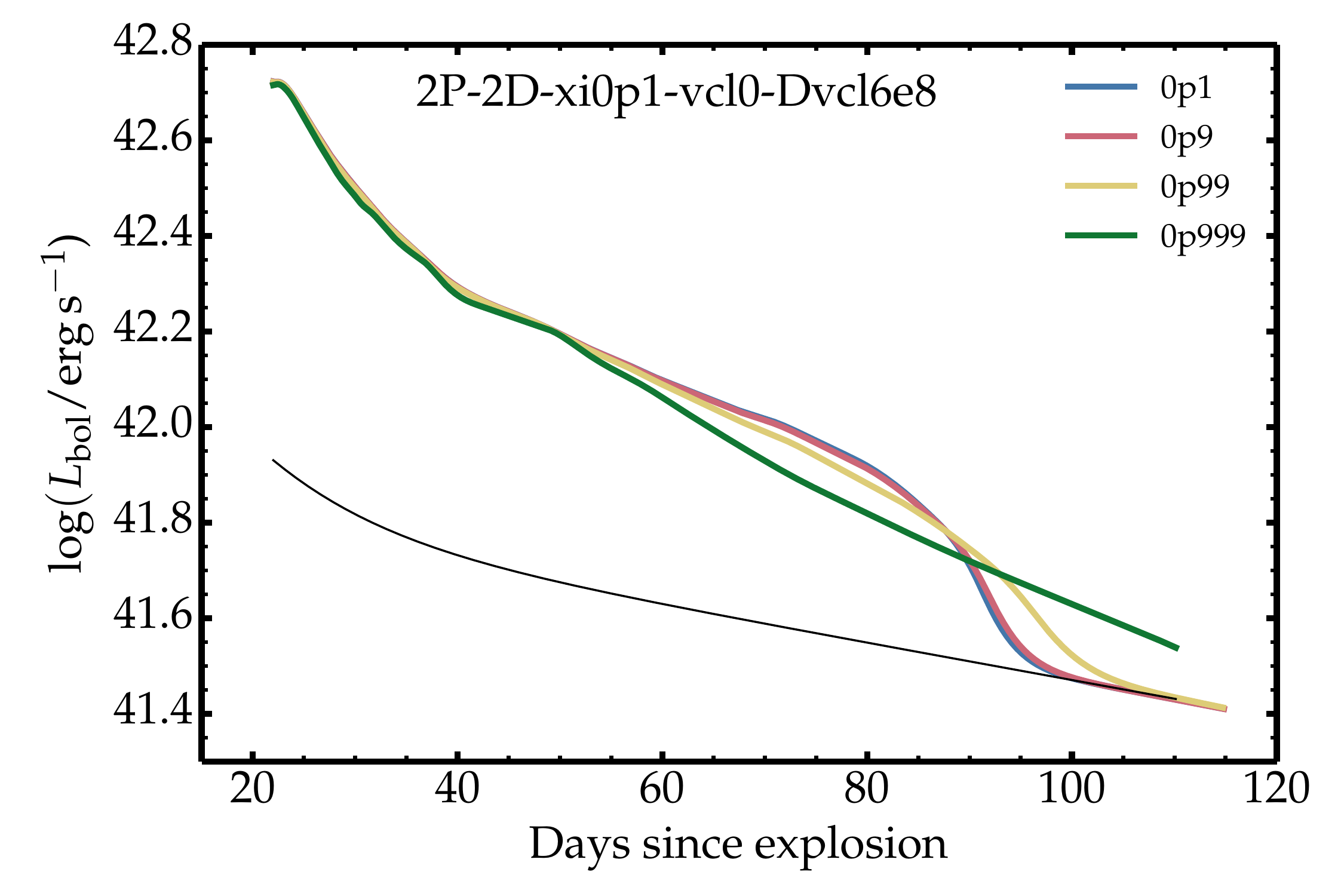}
\end{center}
\vspace{-0.5cm}
   \caption{Left: Bolometric light curves for 12\,\msun\ 2P smooth 1D ejecta but adopting a fixed albedo of 0.1 (absorption dominates the opacity), 0.9, 0.99, and 0.999 (scattering vastly dominates the opacity).  Right: Same as left, but now for a 2P-2D clumped ejecta with $\xi_0=$\,0.1, $V_{\rm cl}=$\,0\,\kms\ and $\Delta V_{\rm cl}=$\,6000\,\kms.}
   \label{fig_lbol_albedo}
\end{figure*}

We have tested the influence of the albedo in our simulations, although retaining for simplicity a uniform value for the whole ejecta (allowing for a meaningful depth dependent albedo as in Fig.~\ref{fig_chi_albedo} would require a nonLTE treatment). The left panel of Figure~\ref{fig_lbol_albedo} shows the bolometric light curves for a set of 12\,\msun\ 2P smooth 1D ejecta. In these simulations, the value of the adopted albedo was varied to cover from an absorption dominated plasma (albedo of 0.1) to a strongly scattering-dominated plasma (albedo of 0.999). The results for an albedo of 0.1 and 0.9 are essentially identical, which, together with the properties shown in Fig.~\ref{fig_chi_albedo}, suggests the results presented in this paper are sound. Interestingly, as the albedo is `unphysically' increased to a value of 0.99 and 0.999, the bolometric light curve stars dipping below the other curves  at about 50\,d, while the photospheric phase is correspondingly extended (each ejecta has the same amount of stored radiation). Our interpretation is that as the scattering opacity is increased, the gas becomes less and less coupled to the radiation, and its emissivity drops, inhibiting its cooling. This weak coupling makes the gas temperature `drift' from the radiation temperature. We find that at 70\,d after explosion, the model with an albedo of 0.999 has a twice higher gas temperature through most of the ejecta relative to the case with an albedo of 0.1. This higher temperature implies a much higher opacity (see Eq.~\ref{eq_kappa}) and diffusion time, which explain the fainter luminosity and the longer phostospheric phase. A similar behavior was seen by \citet{KW09} in their simulation of Type II SN ejecta  using an artificially enhanced electron-scattering opacity. They attributed it to the larger opacity and thus larger optical depth of the ejecta. In our simulation, the opacity is unchanged but the effective opacity is increased because of the shift to a higher gas temperature. This effect probably occurs too in the simulation of \citet{KW09}.

   We have conducted the same experiment but this time in 2D using a clumped ejecta (right panel of Figure~\ref{fig_lbol_albedo}). We find that the albedo has the same effect as in the 1D simulations. This effect is negligible for an albedo below 0.9, and all 2D clumped ejecta yield a similar light curve. The  effect is strong for an albedo greater than 0.9, and dominates over the influence of clumping. For a strongly scattering dominated plasma (which is quite unphysical; see Fig.~\ref{fig_chi_albedo}), the photosphere is pushed far out in the ejecta, in layers where the adopted clumping is weak. Hence, the effect of clumping is dwarfed for a high albedo.

\end{document}